# Calculation of *p*-values for quadratic statistics in pulsar timing arrays

Rutger van Haasteren [*]

*Max-Planck-Institut für Gravitationsphysik (Albert-Einstein-Institut), Callinstraße 38, D-30167 Hannover, Germany Leibniz Universität Hannover, D-30167 Hannover, Germany*

Pulsar Timing Array (PTA) projects have reported various lines of evidence suggesting the presence of a stochastic gravitational wave (GW) background in their data. One key line of evidence involves a detection statistic sensitive to interpulsar correlations, such as those induced by GWs. A $p$-value is then calculated to assess how unlikely it is for the observed signal to arise under the null hypothesis $H_0$, purely by chance. However, PTAs cannot empirically draw samples from $H_0$. As a workaround, various techniques are used in the literature to approximate $p$-values under $H_0$. One such technique, which has been characterized as a model-independent method, is the use of "scrambling" transformations that modify the data to cancel out pulsar correlations, thereby simulating realizations from $H_0$. In this work, scrambling methods and the detection statistic are investigated from first principles. The $p$-value methodology that is discussed is general, but the discussions regarding a specific detection statistic apply to the detection of a stochastic background of gravitational waves with PTAs. All methods in the literature to calculate $p$-values for such a detection statistic are rigorously analyzed, and analytical expressions are derived for the distribution of the detection statistic and the corresponding $p$-values. All this leads to the conclusion that scrambling methods are *not* model independent and thus not completely empirical. With a single realization of data our results are necessarily always model dependent, which any analysis will need to accept. Instead of scrambling approaches, rigorous Bayesian and Frequentist $p$-value calculation methods are advocated, the evaluation of which depend on the generalized $\chi^2$ distribution. This view is consistent with the posterior predictive $p$-value approach that is already in the literature. Efficient expressions are derived to evaluate the generalized $\chi^2$ distribution of the detection statistic on real data. It is highlighted that no Frequentist $p$-values have been calculated correctly in the PTA literature to date.

## I. INTRODUCTION

Recently, four pulsar timing array (PTA) collaborations reported evidence for a low-frequency signal that is correlated among pulsars. Such a signal is what is expected from a stochastic background of gravitational waves (GWB [1–4]) generated by a population of supermassive black-hole binaries at the centers of galaxies [5–7], or more exotic sources [8]. Initial detection claims of a gravitational wave (GW) signal are made in accordance with an established detection checklist [9], which prescribes multiple detection requirements.

One requirement described by the detection protocols is the calculation of a detection significance or $p$-value under the null hypothesis $H_0$. One method to estimate the detection significance recommended in the checklist is through the use of bootstrap methods such as phase scrambling (see section 2.1 of [9]). In such a scrambling method, the complex phases of the Fourier-based data are randomized, thereby maintaining all noise properties but negating all interpulsar correlations. The intuition behind this method is that this gives an empirical estimate of the background distribution of the detection statistic without making any assumptions regarding the noise model.

The PTA community has done extensive simulation work to demonstrate that bootstrap methods provide reasonable estimates of detection significance. However, a theoretical proof of the reliability of these estimates has yet to be established. In this paper, we take a formal approach and show that these "scrambling" methods can be derived from first principles. Along the way, we derive various properties of the background estimates and the corresponding $p$-values obtained using these methods. We outline the connection between scrambling methods and drawing realizations of data straight from $H_0$, demonstrating that scrambling methods also depend on the $H_0$ model and its parameters, similar to other methods for calculating $p$-values.

[*]Contact author: rutger@vhaasteren.com

Moreover, we show that scrambling is mathematically equivalent to $p$-value calculation with the added assumption that we knew the complex amplitudes prior to the analysis. This reliance on prior knowledge of model parameters and data amplitudes makes them vulnerable to model misspecification, and we argue that their primary motivation in the literature may be overstated.

Instead, in this paper, we arrive at Bayesian and Frequentist $p$-value calculation procedures that numerically leverage the generalized $\chi^2$ distribution that was explored in [10]. The Bayesian version of this is equal to the posterior predictive $p$-value from [11]. The Frequentist $p$-value we advocate for has not been explored in detail in the literature.

In Sec. II we give an overview of the contents of this paper, including notation and how to effectively read it.

## II. HOW TO READ THIS PAPER AND NOTATION

The objective of this manuscript is to discuss the detection statistic and $p$-values for GWB searches in PTA data from first principles. In order to do this rigorously, some sections in this paper get dense. However, this paper does not need to be read in order, and some parts can be understood without reading all previous sections. We therefore provide a guide here on how to read the paper. We also provide a summary of notation that is used throughout this work.

### A. Reading order

Overall, Sec. III is necessary as background for the rest of the paper. After reading Sec. III, the reader can choose to read:

(i) Sections IV–VI to understand scrambling procedures. Section VIII then summarizes everything with an overview.

(ii) Section VII to understand the posterior predictive $p$-values. Section VIII then summarizes everything with an overview.

(iii) Section IX to understand how to efficiently implement the generalized $\chi^2$ distribution in real PTA analysis codes

(iv) Dive deep into the Appendixes to appreciate the mathematical details that underpin some of the calculations in the paper

### B. Outline

In Sec. III we give a brief review of null-hypothesis testing in PTAs, including a formal derivation of the optimal detection statistic, and a simple toy model that we use in most of the paper is introduced. We also discuss the difference between the optimal statistic in the literature and Neyman-Pearson optimal statistics. Then, in Sec. IV we derive scrambling methods from first principles. The spread of the detection statistic is then calculated analytically in Sec. V, and in Sec. VI we phrase all the above in terms of polar coordinates where we can appropriately interpret scrambling methods. In Sec. VII we introduce the concept of varying model parameters, and we describe how all distributions relate to one another in Sec. VIII. In Sec. IX we present an efficient formalism to calculate the generalized $\chi^2$ distribution on real PTA data, after which we provide some concluding remarks in Sec. X.

TABLE I. Notation used throughout this paper.

| Symbol | Description |
|---|---|
| $x$, $y$ | Real-valued data |
| $z$ | Complex-valued data |
| $z^\dagger$, $z_j^*$ | Hermitian conjugate and complex conjugate |
| $J$ | Base imaginary number with $J^2 = -1$ |
| $H_0$ | Null hypothesis |
| $H_1$ | Alternative hypothesis (the complement of $H_0$) |
| $H_S$ | Signal hypothesis |
| $M$ | Number of pulsars/detectors (no subscript) |
| $\mathbf{A}$ | A matrix with elements $A_{ab}$ |
| $\mathbf{N}$ | Data covariance under $H_0$ |
| $\mathbf{C}$ | Data covariance under $H_S$ |
| $x \sim \mathcal{N}(0, \mathbf{\Sigma})$ | $x$ is normally distributed (mean 0 and covariance $\mathbf{\Sigma}$) |
| $z \sim \mathcal{N}^{\mathbb{C}}(0, \mathbf{\Sigma})$ | $z = \frac{x+Jy}{\sqrt{2}}$ with $x, y \sim \mathcal{N}(0, \mathbf{\Sigma})$ |
| $\langle\cdot\rangle_G$ | Ensemble average over the ensemble $G$ |
| $\mathbb{E}_G[\cdot]$ | Same as $\langle\cdot\rangle_G$, for extra clarity in complex expressions |
| $\mathrm{d}\nu(S)$ | Haar measure (uniform probability distribution over $S$) |
| $\mathrm{Tr}[\mathbf{X}]$ | Trace of matrix $\mathbf{X}$ |
| $\mathrm{Gamma}(a, b)$ | Gamma distribution with shape $a$ and scale $b$ |
| $P(x < a)$ | Probability that $x < a$ |
| $p(x\|a)$ | Probability density of $x$ conditioned on $a$ |
| $p$ | $p$-value (no parentheses) |
| $\tilde{\mathbf{Q}}$, $\tilde{z}$ | Noise-weighted matrices and vectors |
| $(x, y)_C$ | Inner product between $x$ and $y$, using weighting $\mathbf{C}$ |
| $\mathrm{Tr}_C[\mathbf{A}, \mathbf{B}]$ | Trace between $\mathbf{A}$ and $\mathbf{B}$, using weighting $\mathbf{C}$ |
| $U \sim \mathcal{H}U(M)$ | $U$ is a Haar-distributed random unitary transformation |
| $\phi \sim \mathcal{H}S^{2M-1}$ | $\phi$ is a Haar-distributed random variable on $S^{2M-1}$ |

### C. Notation

In Table I we list the notation we use in this manuscript.

## III. OPTIMAL DETECTION STATISTIC REVIEW

We start off with a review of what has been done regarding detection statistics in the PTA literature: we cover the basic theory, and we introduce a toy model that will help us understand what is happening. Next we also introduce alternate ways to derive and interpret an optimal detection statistic for an isotropic background of gravitational waves, which provides insight into the requirements for the null and signal hypotheses. We discuss how the "optimal" statistic in the PTA literature is actually not optimal in a Neyman-Pearson sense, and we review analytical methods to calculate the detection statistic background distribution.

### A. Toy model

Although we attempt to stay as general as possible in our investigations, some arguments are more easily addressed when considering a specific model. We therefore introduce a model that is inspired by the model that was recently used in van Haasteren ([12], Sec. II): we have $M = 67$ pulsars with locations roughly similar to the locations of the pulsars in the NANOGrav 15 year dataset [13], and each pulsar yields $N_o = 2$ degrees of freedom from a single frequency bin. These two degrees of freedom are represented by a single complex number with independent real and imaginary parts. Generalization of techniques in this manuscript are trivially adapted to realistic applications with more frequency bins—we show in Sec. IX how to do that. In using distributions for complex numbers, we use the following notation:

$$z \sim \mathcal{N}^{\mathbb{C}}(0, \Sigma), \tag{1}$$

$$z_a = \frac{x_a + J y_a}{\sqrt{2}}, \tag{2}$$

$$x \sim \mathcal{N}(0, \Sigma), \tag{3}$$

$$y \sim \mathcal{N}(0, \Sigma) \tag{4}$$

where we use $J$ as the imaginary base number with $J^2 = -1$. The index $a$ represents the pulsar and runs between $[1, M]$. With $\mathcal{N}^{\mathbb{C}}$ we mean a complex random variable where both the real and imaginary parts are independent random draws from the multivariate Gaussian $\mathcal{N}(0, \Sigma)$. We review all this in Appendix A, where we give some useful identities.

In our simplified model, there are only two processes that give a response in these pulsar data: the white noise process $Z_n$ and the signal process $Z_h$. The process $Z_n$ produces independent identically distributed (IID) data with a known or unknown amplitude in each pulsar,

$$n_a \sim \mathcal{N}^{\mathbb{C}}(0, \sigma_a^2), \tag{5}$$

where $n_a$ is the sample of the noise at pulsar $a$, where we use the convention that $a$ labels pulsar $a \in [1, M]$, and $\sigma_a$ is the standard deviation of the noise.

The signal process $Z_h$ produces realizations that are correlated between pulsars. In a single pulsar, realizations of $Z_h$ can be modeled as IID and are therefore indistinguishable from $Z_n$. When analyzing data from multiple pulsars simultaneously the two processes can be distinguished, but not when processing single pulsar data in isolation. A main focus of PTA science are Hellings & Downs (H&D) correlations, which represent the average correlations that an isotropically unpolarized ensemble of gravitational-wave sources would induce. These correlations, which we denote as $\mu(\gamma_{ab})$, depend only on the angular separation $\gamma_{ab}$ of pulsar $a$ and pulsar $b$,

$$\begin{aligned} \xi_{ab} &= \frac{(1 - \cos\gamma_{ab})}{2}, \\ \mu(\gamma_{ab}) &= \frac{3\xi_{ab}\log\xi_{ab}}{2} - \frac{1}{4}\xi_{ab} + \frac{1}{2}(1 + \delta_{ab}), \end{aligned} \tag{6}$$

where $\delta_{ab}$ is the Kronecker delta. We use $\mu_{ab} = \mu(\gamma_{ab})$ for convenience going forward. The realizations of $Z_h$ in the data can now be written as,

$$\begin{aligned} h &\sim \mathcal{N}^{\mathbb{C}}(0, \mathbf{\Sigma}) \\ \Sigma_{ab} &= \mathscr{h}^2 \mu_{ab}. \end{aligned} \tag{7}$$

The parameter $\mathscr{h}$ represents the signal amplitude, which has an interpretation similar to $\sigma_a$ for the noise, and $\mathbf{\Sigma}$ is the covariance of the multivariate Gaussian distributed variable $h$. The full data $z$ can be written as,

$$z_a = h_a + n_a, \tag{8}$$

$$z \sim \mathcal{N}^{\mathbb{C}}(0, \mathbf{C}), \tag{9}$$

$$C_{ab} = \hbar^2 \mu_{ab} + \sigma_a^2 \delta_{ab}. \tag{10}$$

### B. Null hypothesis testing

In classical null hypothesis testing [14], a *test statistic* is designed as a function $D(z)$ of the data $z$, which is used to assess the presence of a certain effect in the data. The null hypothesis $H_0$ represents a model of the data under the assumption that the effect is absent. The null hypothesis is rejected when the observed value of the test statistic, $D_{\mathrm{obs}} \coloneqq D(z_{\mathrm{obs}})$, lies in the extreme tails of the null distribution,

$$P(D > D_{\mathrm{obs}} | H_0) = \int_{D(z)>D_{\mathrm{obs}}} \mathrm{d}z\, p(z|H_0). \tag{11}$$

In other words, we reject $H_0$ when it is highly improbable that $H_0$ could produce data resulting in a detection statistic as extreme as $D_{\mathrm{obs}}$. The probability $p = P(D > D_{\mathrm{obs}}|H_0)$, known as the one-sided $p$-value, represents the probability that $H_0$ could generate any data $z$ that results in a detection statistic at least as large as $D_{\mathrm{obs}}$. Importantly, the $p$-value does *not* represent the probability that $H_0$ is true, given the data. The threshold $p$-value $p_t$, below which $H_0$ is rejected, is a matter of statistical convention and has been the subject of considerable debate in the scientific literature [15–18].

In the PTA literature, null hypothesis testing is often encountered in studies analyzing the PTA data in search of a stochastic background of GWB. The detection statistic [19–23] is crafted to specifically respond to correlations in the data between pulsars in a specific pattern unique to GWs. In order to assess how significant an observed value of a detection statistic $D_{\mathrm{obs}}$ is, we need to have a way to calculate the distribution of $D(z)$ when the data $z \sim p(z|H_0)$. Methods for obtaining that distribution are the topic of this manuscript.

### C. Deriving the detection statistic

Detection statistics for PTAs that focus on the correlations between pulsars have been around since Jenet *et al.* [19] first presented their methodology. Their treatment was a significant advancement, but nonetheless still based on an oversimplification of the PTA signal and noise model. Subsequent work has improved the sensitivity of the detection statistic, the new version of which is commonly referred to as the "optimal statistic." We warn the reader that this is just a name that has been adopted in the literature, as we show in later sections that this statistic is not the statistic that maximizes the detection probability for a fixed false alarm rate. In this manuscript, most results hold for detection statistics defined by quadratic filter $\mathbf{Q}$ subject to certain conditions, without assuming any sort of optimality.

The detection statistic for an isotropic GWB in PTAs that is currently used in the literature is a quadratic function of the data, which can be written as [20–23],

$$D(z, \theta) = \sum_{ab} Q_{ab} z_a^* z_b = z^\dagger \mathbf{Q} z, \tag{12}$$

where $D(z, \theta)$ is our detection statistic, $\mathbf{Q} = \mathbf{Q}(\theta)$ is a filter matrix that depends on the model parameters $\theta$, and the data $z$ is a vector of (complex) numbers as described in Sec. III A. Typically $\mathbf{Q}$ is defined to be a matrix with zeros on the diagonal. This forces the filter to only use the cross-correlations, and not the autocorrelations. The quadratic filter is then carefully chosen to maximize some figure of merit under the signal hypothesis $H_S$. We do not write $H_1$ for the signal hypothesis, because $H_S$ represents some specific signal model, whereas $H_1$ refers to anything that is not $H_0$.

In our derivation below we find that under a certain choice of null-hypothesis (the so-called CURN model, which we define later) the quadratic filter $\mathbf{Q}$ will automatically have zeros on the diagonal even if we do not put in that constraint. In general a detection statistic does not need to be quadratic in the data (and we will see later in this paper that the PTA optimal statistic actually is *not* quadratic in the data in any method that is used in practice, because we need to first determine the model parameters), but quadratic estimators have attractive properties. Under certain conditions, it can be shown that quadratic estimators include the uniformly most powerful test [24,25].

In what follows we derive the detection statistic as a quadratic filter under the assumption that the data obeys

$$H_0\colon\ \langle z z^\dagger \rangle_{H_0} = \mathbf{N}, \tag{13}$$

$$H_S\colon\ \langle z z^\dagger \rangle_{H_S} = \mathbf{C}, \tag{14}$$

with the understanding that the noise covariance is given by $\mathbf{N}$ and that the signal hypothesis $H_S$ introduces extra or different correlations. We parametrize the $\mathbf{N}$ and $\mathbf{C}$ matrices with parameters $\theta$. However, in the below derivations we assume that we know $\theta$ exactly, which makes the hypotheses "simple hypotheses" (not dependent on free parameters). We define the "signal-to-noise" functional, also called the *deflection*, as

$$L(z) \coloneqq \frac{D(z,\theta) - \langle D(z,\theta) \rangle_{H_0}}{\sqrt{\langle D(z,\theta)^2 \rangle_{H_0} - \langle D(z,\theta) \rangle_{H_0}^2}}, \tag{15}$$

and we impose the normalization constraint on the detection statistic

$$\langle D(z)^2 \rangle_{H_0} - \langle D(z) \rangle_{H_0}^2 = 1. \tag{16}$$

Under that constraint, the expression for $L$ simplifies,

$$L_c(z) = D(z,\theta) - \langle D(z,\theta)\rangle_{H_0}, \tag{17}$$

where we used the suffix $c$ to remember it is the constrained version of $L$. The deflection $L$ has been used in the statistics literature for a long time, and it can be justified by many arguments. It has an interpretation as a signal to noise ratio. However, it is not possible to prove that for a given false alarm probability an increase in deflection will result in an increase in detection probability. The physical meaning of the deflection as a detection criterion is not obvious. Regardless, it is often a useful criterion *in practice*, and it has been useful for PTA purposes: most “optimal statistic” derivations to date turn out to come from optimizing the deflection under $H_S$.

Below we outline two alternate derivations of the optimal filter $\mathbf{Q}$ that maximizes $L$ in expectation under $H_S$.

#### *1. Method 1: Derivation via the Cauchy-Schwarz inequality*

A standard approach in the literature is to cast the problem in terms of an inner product. With our notation, we define the inner product on the space of matrices. In our case, we define

$$(\mathbf{A},\mathbf{B}) \coloneqq \mathrm{Tr}[\mathbf{NANB}], \tag{18}$$

for any matrices $\mathbf{A}$ and $\mathbf{B}$. Because $\mathbf{N}$ is positive definite, this definition satisfies the requirements of an inner product. We start with the average of the detection statistic under the two hypotheses

$$\langle D(z,\theta)\rangle_{H_0} = \mathrm{Tr}[\mathbf{QN}], \tag{19}$$

$$\langle D(z,\theta)\rangle_{H_S} = \mathrm{Tr}[\mathbf{QC}], \tag{20}$$

so that the difference is

$$\langle L_c\rangle_{H_S} \coloneqq \langle D(z,\theta)\rangle_{H_S} - \langle D(z,\theta)\rangle_{H_0} = \mathrm{Tr}[\mathbf{Q}(\mathbf{C}-\mathbf{N})]. \tag{21}$$

It is straightforward to show that this difference may be written using our custom inner product as

$$\langle L_c\rangle_{H_S} = (\mathbf{N}^{-1}\mathbf{C}\mathbf{N}^{-1} - \mathbf{N}^{-1},\mathbf{Q}), \tag{22}$$

while the variance in the detection statistic can be expressed as

$$\Delta^2 \coloneqq \langle D(z,\theta)^2\rangle_{H_0} - \langle D(z,\theta)\rangle_{H_0}^2 = (\mathbf{Q},\mathbf{Q}). \tag{23}$$

Hence, the signal-to-noise ratio function $L$ can be written as

$$\langle L\rangle_{H_S} = \frac{(\mathbf{N}^{-1}\mathbf{C}\mathbf{N}^{-1} - \mathbf{N}^{-1},\mathbf{Q})}{\sqrt{(\mathbf{Q},\mathbf{Q})}}. \tag{24}$$

The Cauchy–Schwarz inequality then implies that the maximum is achieved when the filter $\mathbf{Q}$ is proportional to

$$\mathbf{Q} \propto \mathbf{N}^{-1}(\mathbf{C}-\mathbf{N})\mathbf{N}^{-1}. \tag{25}$$

The proportionality constant is determined by enforcing the normalization condition of Eq. (16) on the variance of the detection statistic, which finally yields,

$$\mathbf{Q} = \frac{\mathbf{N}^{-1}(\mathbf{C}-\mathbf{N})\mathbf{N}^{-1}}{\mathrm{Tr}[(\mathbf{N}^{-1}(\mathbf{C}-\mathbf{N}))^2]^{1/2}}. \tag{26}$$

#### *2. Method 2: Derivation via a Lagrange multiplier*

An alternative derivation is obtained by “whitening” the data. We define the following transformed matrices:

$$\mathbf{B} \coloneqq \mathbf{N}^{-1/2}\mathbf{C}\mathbf{N}^{-1/2}, \tag{27}$$

$$\mathbf{A} \coloneqq \mathbf{I} - \mathbf{B}, \tag{28}$$

$$\mathbf{X} \coloneqq \mathbf{N}^{1/2}\mathbf{Q}\mathbf{N}^{1/2}. \tag{29}$$

In these variables the normalization constraint of Eq. (16) takes the simple form

$$\mathrm{Tr}[\mathbf{X}^2] = 1, \tag{30}$$

and the expected $L_c$ under $H_S$ can be written as

$$\langle L_c\rangle_{H_S} = \langle D(z,\theta)\rangle_{H_S} - \langle D(z,\theta)\rangle_{H_0} = \mathrm{Tr}[\mathbf{AX}]. \tag{31}$$

Our goal is to maximize

$$\langle L_c\rangle_{H_S} = \mathrm{Tr}[\mathbf{AX}], \tag{32}$$

subject to the constraint $\mathrm{Tr}[\mathbf{X}^2] = 1$. Introducing a Lagrange multiplier $\mu$, we consider the Lagrangian

$$\mathcal{L}(\mathbf{X},\mu) = \mathrm{Tr}[\mathbf{AX}] + \mu(\mathrm{Tr}[\mathbf{X}^2] - 1). \tag{33}$$

Taking the derivative with respect to $X$ and setting it equal to zero gives

$$\mathbf{A} + 2\mu\mathbf{X} = 0, \tag{34}$$

or equivalently,

$$\mathbf{X} = -\frac{1}{2\mu}\mathbf{A}. \tag{35}$$

Notice how $\mathbf{A}$ and $\mathbf{X}$ turn out to commute. The normalization condition then immediately determines

$$\mu = -\frac{1}{2}\sqrt{\mathrm{Tr}[\mathbf{A}^2]}. \tag{36}$$

Returning to the original variable $\mathbf{Q}$ via

$$\mathbf{Q} = \mathbf{N}^{-1/2}\mathbf{X}\mathbf{N}^{-1/2}, \tag{37}$$

we obtain an expression for the optimal filter that is equivalent to the result from Method 1,

$$\mathbf{Q} = \frac{\mathbf{N}^{-1}(\mathbf{C}-\mathbf{N})\mathbf{N}^{-1}}{\mathrm{Tr}[(\mathbf{N}^{-1}(\mathbf{C}-\mathbf{N}))^2]^{1/2}}. \tag{38}$$

Both derivations lead to the same optimal form for $\mathbf{Q}$ that maximizes the deflection under $H_S$. The method with the Lagrange multipliers is flexible in the sense that extra constraints may be introduced in the derivation. Although playing around with other constraints is instructive, in the end we have found no practical use for it. Notice how the numerator of $\mathbf{Q}$ contains the term $\mathbf{C}-\mathbf{N}$. Interestingly, if the null hypothesis $H_0$ is the so-called common uncorrelated red noise (CURN) model, we have $N_{ab} = \delta_{ab}C_{ab}$. In that case, the diagonal components of $\mathbf{Q}$ vanish. So, even though we made no restriction on the use of the autocorrelations, under the CURN null hypothesis the autocorrelations are not used in the optimal detection statistic for a GWB.

This form of the detection statistic more explicitly treats the $H_0$ and $H_S$ hypotheses than other presentations of the optimal detection statistic in the literature [20–23,26,27]. Our formulation works for *any* two hypotheses, and we do not require the null hypothesis to represent datasets that have no correlations between pulsars. However, what we derive here is fully consistent with other forms in the literature.

### D. The Neyman-Pearson optimal statistic

Even though the detection statistic defined by the quadratic filter of Eq. (38) is referred to in the literature as the "optimal statistic," it is not the uniformly most powerful (UMP) test as can be derived with the Neyman-Pearson Lemma [28] for simple (noncomposite) hypotheses. If we define $\mu_S = \langle D(z)\rangle_{H_S}$, $\sigma_0^2 = \langle D(z)^2\rangle_{H_0}$, and $\sigma_S^2 = \langle D(z)^2\rangle_{H_S}$, then the optimal statistic maximizes $\mu_S/\sigma_0$, so it maximizes the signal to noise ratio. In some references, such as Rosado *et al.* [29], an alternative statistic is introduced that is more robust in the stronger-signal regime: the statistic that maximizes $\mu_S/\sigma_S$. While this is true, neither of the tests one obtains from maximizing $\mu_S/\sigma_0$ or $\mu_S/\sigma_S$ is a statistic that is optimal in the Neyman-Pearson sense.

Through the Neyman-Pearson Lemma we can find the UMP with the likelihood ratio $\Lambda = p(z|\theta, H_S)/p(z|\theta, H_0)$. For a model with known model parameters $\theta$ we can leave out the normalizations of the Gaussian likelihoods, and immediately find after taking the logarithm,

$$\mathbf{Q} = \frac{\mathbf{N}^{-1}-\mathbf{C}^{-1}}{\mathrm{Tr}[(\mathbf{C}^{-1}(\mathbf{C}-\mathbf{N}))^2]^{1/2}} \tag{39}$$

$$= \frac{\mathbf{N}^{-1}(\mathbf{C}-\mathbf{N})\mathbf{C}^{-1}}{\mathrm{Tr}[(\mathbf{C}^{-1}(\mathbf{C}-\mathbf{N}))^2]^{1/2}}. \tag{40}$$

We see that there is a lot of similarity between the Neyman-Pearson optimal statistic for a GWB, and the optimal deflection detection statistic from Eq. (38). We repeat that the version of the detection statistic that is currently in the literature is the statistic one gets when optimizing the deflection under $H_S$, with the added requirement that $H_0$ is assumed to be characterized by $N_{ab} = \delta_{ab}C_{ab}$. This additional requirement assures that only cross-correlations between pulsars are used. In the weak-signal regime, the Neyman-Pearson optimal statistic becomes equal to the optimal deflection detection statistic from Eq. (38).

For the Neyman-Pearson optimal statistic we cannot use the likelihood ratio while ignoring the autocorrelations in a consistent way. For the remainder of this manuscript we will assume that $\mathbf{Q}$ was constructed through Eq. (38). For an investigation of the Neyman-Pearson optimal detection statistic that only uses cross-correlations, we refer to van Haasteren *et al.* [30].

### E. Analytical background estimate

If the model parameters are known exactly, Hazboun *et al.* [10] pointed out that the detection statistic of Eq. (12) is a quadratic combination of normal variables. Such a random variable follows a generalized $\chi^2$ distribution, and $p$-values can be calculated analytically under the assumption of $H_0$. In the context of the toy model of Sec. III A, if $\mathbf{\Sigma}$ is known, we can use the Cholesky decomposition $\mathbf{L}$ of the model covariance $\mathbf{\Sigma} = \mathbf{L}\mathbf{L}^T$ to write the detection statistic in terms of zero-mean unit-variance random variables $\xi_a \sim \mathcal{N}^{\mathbb{C}}(0,1)$. In fact, in our toy model $L_{ab} = \delta_{ab}\sqrt{\sigma_a^2 + \hbar^2}$ under $H_0$. This is equivalent to the CURN model in the PTA literature. In other words, $\mathbf{\Sigma}$ is diagonal under $H_0$. However, in general we would use a dense Cholesky factor $\mathbf{L}$ or some other matrix square root,

$$D(z,\theta) = \xi^\dagger \mathbf{L}^T\mathbf{Q}\mathbf{L}\xi \tag{41}$$

$$= \xi^\dagger \mathbf{U}\mathbf{\Lambda}\mathbf{U}^T\xi \tag{42}$$

$$= \tilde{z}^\dagger \mathbf{\Lambda}\tilde{z} = \sum_a \lambda_a |\tilde{z}_a|^2 \tag{43}$$

where $\mathbf{U}$ is an orthogonal matrix with as columns the eigenvectors of $\mathbf{L}^T\mathbf{Q}\mathbf{L}$, and $\mathbf{\Lambda}$ is the diagonal matrix with eigenvalues $\lambda_a = \Lambda_{aa}$ on the diagonal. Since $\mathbf{U}$ is unitary, the elements of $\tilde{z}$ are also distributed as $\tilde{z}_a \sim \mathcal{N}^{\mathbb{C}}(0,1)$. Each component of $|\tilde{z}_a|^2$ in Eq. (43) represents a $\chi^2$ distribution with two degrees of freedom.

With Eq. (41) we can then follow Hazboun *et al.* [10], and describe the distribution of $D(z)$ as a generalized chi-square distribution with weights $w$ using the Imhof approximation [31,32]. The only thing we need to remember is that we have complex random variables, and therefore each eigenvalue $\lambda_a$ is duplicated (appears twice), representing two random variables for each $\lambda_a$. This approach generalizes easily to multiple frequency bins in realistic PTAs, as we just need to combine the weights from all frequencies into a single weights vector.

The analytical background estimate from Hazboun *et al.* [10] has seen somewhat limited adoption in real analyses, because it is framed in the time-domain. Therefore, the eigenvalue decomposition is impractical in contemporary datasets due to the matrix sizes involved. In Sec. IX we show how to use rank-reduced methods to efficiently calculate the background distribution.

## IV. SCRAMBLING BACKGROUND ESTIMATION

The data of PTA projects contain only a single time-series of signal/noise, spanning over a decade for most pulsars, where the noise description of each pulsar is unique. There is some understanding of some noise contributions to the data, but in general the community consensus is that the noise model can be improved with additional modeling effort and observations. Our effective null hypothesis $H_0$ *may not represent reality*, which reduces our ability to confidently quantify the significance of a detection.

Inspired by the use of time-slides in the LIGO literature, the PTA community sought to use the PTA data itself to construct an estimate of the detection statistic background distribution under the null hypothesis. Two methods were suggested as viable replacements of $p(z|H_0)$: sky scrambling [33] and phase scrambling [34]. In sky scrambling, the locations of the pulsars in the (GW) model are artificially changed such that the detection statistic $D(z)$ of Eq. (12) is no longer sensitive to the correlations in the data. In phase scrambling or phase shifting, the correlations in the data are negated by artificially introducing complex phase rotations that negate the interpulsar correlations in the data.

The proposed scrambling methods have seen near-universal adoption in the PTA community, and the resulting $p$-values seem reasonable [1] in the literature. Sky scrambling is slightly less prominently featured in published results because the more complicated dependence of the data under pulsar sky location changes makes the resulting $p$-values less well understood, and some have critiqued the dependence of different sky scrambles [35]. However, both scrambling methods are widely used. In this section we derive scrambling methods from first principles and give them a firm theoretical basis.[1]

### A. Weighted inner product

We take a formal approach to scrambling, which we define in terms of a transformation of the data $z' = S(z)$. In order to proceed, we first introduce some notation regarding a new inner product that will help us with our derivation. This notation is motivated by the observation that we can rewrite the definition of our models and the detection statistic more elegantly.

Remember that we defined $H_0$ as a multivariate normal distribution with $z \sim \mathcal{N}^{\mathbb{C}}(0, \mathbf{N})$, where the relationship between $\mathbf{N}$ under $H_0$ and $\mathbf{C}$ is typically $N_{ab} = \delta_{ab} C_{ab}$. For scrambling procedures that is required for $H_0$. Two equivalent and sufficient descriptions of $H_0$ are to state that $z$ is a multivariate normally distributed variable with,

$$\mathbb{E}[z_a z_b^*] = N_{ab}, \tag{44}$$

$$\mathbb{E}[\mathbf{N}^{-1} z z^\dagger] = \mathbf{I} \tag{45}$$

where we note that $z$ is a complex random variable. More in line with other forms in the literature, the detection statistic we derived in Sec. III C can be written as,

$$D(z, \theta) = z^\dagger \mathbf{Q} z, \tag{46}$$

$$W_{ab} = (1 - \delta_{ab}) \mu(\gamma_{ab}), \tag{47}$$

$$\mathbf{Q} = \frac{\mathbf{N}^{-1} \mathbf{W} \mathbf{N}^{-1}}{(\mathrm{Tr}(\mathbf{N}^{-1} \mathbf{W} \mathbf{N}^{-1} \mathbf{W}))^{1/2}}. \tag{48}$$

Compared to other expressions in the literature, that means we sum over all pulsar combinations $(a, b)$ and *not* just unique pairs. While other derivations would then pick up a factor of two in the denominator, we do not because we work with complex random variables. For real-valued data, Eq. (47) will have a factor of two multiplying the trace. We can make things look nicer if we define a new inner product on vector $x$ and $y$. Beware that this is a different inner product than we used in Sec. III C 1. We write,

[1]We acknowledge that both sky scrambling and phase scrambling methods were introduced with the idea of calibrating Bayes Factors [33,34], rather than obtaining $p$-values the way we do in this paper. However, their intended use is the same as here: to emulate $z \sim p(z|H_0)$. We believe there is enough similarity to make many of the arguments carry over to the use of scrambling methods for Bayes Factor calibration. Additionally, subsequent papers have used scrambling methods for $p$-value calculations with fixed model parameters (e.g., [36]). So, while hedging that we position scrambling methods not as intended in their inception, we believe it is most insightful to frame them this way for the purposes of this paper.

$$(x, y)_N \coloneqq x^\dagger \mathbf{N}^{-1} y. \tag{49}$$

Because of the deep connection between traces and inner products, we similarly need to define a new trace on matrix $\mathbf{A}$ and $\mathbf{B}$, written with commas,

$$\mathrm{Tr}_N[\mathbf{A}, \mathbf{B}] \coloneqq \mathrm{Tr}(\mathbf{N}^{-1}\mathbf{A}\mathbf{N}^{-1}\mathbf{B}). \tag{50}$$

The Cartesian trace is still written without commas. The inner product can also incorporate multiples,

$$(x, \mathbf{A}, y)_N = x^\dagger \mathbf{N}^{-1}\mathbf{A}\mathbf{N}^{-1} y. \tag{51}$$

This new inner product allows an interpretation as a noise-weighting transformation of our vectors and matrices: $\tilde{x} = \mathbf{N}^{-1/2}x$ and $\tilde{\mathbf{A}} = \mathbf{N}^{-1/2}\mathbf{A}\mathbf{N}^{-1/2}$. Under the new inner product, we can write the optimal statistic as,

$$D(z, \theta) = \frac{(z, \mathbf{W}, z)_N}{(\mathrm{Tr}_N[\mathbf{W}, \mathbf{W}])^{1/2}} = \alpha(z, \mathbf{W}, z)_N, \tag{52}$$

where we defined the normalization constant $\alpha = (\mathrm{Tr}_N[\mathbf{W}, \mathbf{W}])^{-1/2}$ as we did in Sec. III C.

### B. Requirement 1: Unitarity

Armed with our newly defined inner product and notation, we are in a position where we can define the requirements of our scrambling operators. As a first requirement, we stated that we would like to find scrambling transformations under which $H_0$ is invariant. The intuition is clear: if $H_0$ is still valid after the transformation, all of our noise analysis is still valid. We are looking for transformations $S$ that transform the data $z' = S(z)$. The collection of all such transformations we denote with $T$. Now we define the collection of transformations $G \subset T$,

$$G = \{S \in T : (S(z), S(z))_N = (z, z)_N \quad \forall z \in \mathbb{C}^M\}. \tag{53}$$

In words, we are looking for transformations of $z$ that do not change the squared norm $(z, z)_N$, as is required by Eq. (44).

At this point we introduce an extra restriction: we limit ourselves to linear operators. Linear operators under which the squared norm is invariant are part of the unitary group $U(M)$. That is, $G = U(M)$. The unitary group is defined with the requirement that if $S \in U(M)$, then $\mathbf{S}^\dagger\mathbf{S} = \mathbf{S}\mathbf{S}^\dagger = \mathbf{I}$. The unitary group $G$ we defined is unitary with respect to our weighted inner product. In matrix form, the elements of of the weighted unitary group $G$, denoted with $S_w$, can be written as: $\mathbf{S}_w = \mathbf{N}^{1/2}\mathbf{S}\mathbf{N}^{-1/2}$, where $S \in U(M)$.

A consequence of unitarity under $G$ is that,

$$\mathbb{E}_{H_0}[z' z'^\dagger] = \mathbb{E}[\mathbf{S}_w z z^\dagger \mathbf{S}_w^\dagger] = \mathbf{N}, \tag{54}$$

where the expectation is taken over $H_0$. This confirms that $H_0$ is invariant under $G$. The interpretation of elements in the group $G$ is: first whiten with $\mathbf{N}^{-1/2}$, then carry out a unitary transformation, then undo the whitening with $\mathbf{N}^{1/2}$.

We now touch on some extra insights we can gain from this invariance of $H_0$ under $G$. The optimal detection statistic $D(z, \theta)$ was formally derived in the literature under the assumption of $H_0$. Therefore, the detection statistic is still the optimal detection statistic on the transformed data. Even though $H_0$ is invariant under $G$, the data is obviously transformed, so the numeric value of $D(z, \theta)$ of course does change under $G$. We can see that the form of the detection statistic is unchanged under $G$ by observing that because $H_0$ is invariant, we have,

$$\mathrm{Tr}_N[\mathbf{S}_w^\dagger \mathbf{W}\mathbf{S}_w, \mathbf{S}_w^\dagger \mathbf{W}\mathbf{S}_w] = \mathrm{Tr}_N[\mathbf{W}, \mathbf{W}]. \tag{55}$$

The denominator of the detection statistic is invariant under the unitary transformations of $G$. This makes our life a lot easier, because it enables us to calculate ensemble statistics analytically. We can consider the denominator as a constant under scrambling, and only transform the data $z$.

### C. Requirement 2: $D(z, \theta) = 0$ in expectation

Now that we have formally defined invariance of $H_0$ under our scrambling operations, we need to see what those transformations do to our detection statistic and what extra requirements we need to place on $G$, if any. The detection statistic has an average of zero under $H_0$. We now need to make sure that the detection statistic has an average of zero under *any* zero-mean model when scrambling with weighted unitary transformations. This is our formal requirement:

$$\mathbb{E}_G[D(z, \theta)] = 0 \tag{56}$$

where $z = S_w(z^{\mathrm{obs}})$ with $S_w \in G$ drawn uniformly from $G$. Let us see whether we can verify whether Eq. (56) is already satisfied if we use some definition of uniformity,

$$\mathbb{E}_G[D(z, \theta)] = \mathbb{E}_G[\alpha(S_w(z), \mathbf{W}, S_w(z))_N] \tag{57}$$

$$= \mathbf{E}_{U(M)}[\alpha \tilde{z}^\dagger \mathbf{S}^\dagger \tilde{\mathbf{W}} \mathbf{S} \tilde{z}] \tag{58}$$

$$= \alpha \tilde{z}^\dagger \mathbf{E}_{U(M)}[\mathbf{S}^\dagger \tilde{\mathbf{W}} \mathbf{S}] \tilde{z} \tag{59}$$

with $S \in U(M)$. We have used $\tilde{z} = \mathbf{N}^{-1/2}z$ and $\tilde{\mathbf{W}} = \mathbf{N}^{-1/2}\mathbf{W}\mathbf{N}^{-1/2}$ as the weighted data and the weighted filter, just like above. Under $H_0$ $\mathbf{N}$ is diagonal, so $\tilde{\mathbf{W}}$ is still a traceless matrix. Also, we have that $\langle \tilde{z}\tilde{z}^\dagger \rangle = \mathbf{I}$.

With the expectation over the regular unitary group $U(M)$, we can leverage the random matrix theory literature, writing

$$\mathbf{E}_{U(M)}[\mathbf{S}^{\dagger}\tilde{\mathbf{W}}\mathbf{S}] = \int \mathrm{d}\nu(S)\mathbf{S}^{\dagger}\tilde{\mathbf{W}}\mathbf{S}, \tag{60}$$

where $\mathrm{d}\nu(S)$ is the Haar measure [37] that defines the probability distribution over the set of unitary matrices. The Haar measure is the unique probability measure that is invariant under left and right multiplication by any unitary matrix. This means that when we sample from the Haar measure, or if we integrate over it, the probability distribution respects this invariance. We proceed by noting that $\tilde{\mathbf{W}}$ is Hermitian, meaning it can be diagonalized by some unitary matrix,

$$\tilde{\mathbf{W}} = \mathbf{V}\tilde{\mathbf{W}}_{\mathrm{d}}\mathbf{V}^{\dagger}, \tag{61}$$

where $\mathbf{V}$ is unitary, and $\tilde{\mathbf{W}}_{\mathrm{d}}$ is diagonal. Since the Haar measure is invariant under left and right multiplication by unitary matrices, we can therefore write,

$$\mathbb{E}_{U(M)}[\mathbf{S}^{\dagger}\tilde{\mathbf{W}}\mathbf{S}] = \int \mathrm{d}\nu(S)\mathbf{S}^{\dagger}\tilde{\mathbf{W}}\mathbf{S} \tag{62}$$

$$= \int \mathrm{d}\nu(S)\mathbf{S}^{\dagger}\tilde{\mathbf{W}}_{\mathrm{d}}\mathbf{S} \tag{63}$$

$$= c\mathbf{I}_M. \tag{64}$$

In the last line we used the symmetry of the group of unitary matrices to deduce that the expectation value should be proportional to the identity matrix. Indeed, the unitary group $U(M)$ is isotropic, and it treats all directions equally. The result must therefore be invariant under permutations and index relabelings. We now use the fact that $\tilde{\mathbf{W}}$ is a traceless matrix,

$$\mathrm{Tr}(\mathbb{E}_{U(M)}[\mathbf{S}^{\dagger}\tilde{\mathbf{W}}\mathbf{S}]) = \int \mathrm{d}\nu(S)\mathrm{Tr}(\tilde{\mathbf{W}}_{\mathrm{d}}) = c\mathrm{Tr}(\mathbf{I}) = 0. \tag{65}$$

We therefore conclude that $c = 0$, and,

$$\mathbb{E}_G[D(z,\theta)] = 0. \tag{66}$$

This means that the weighted unitary group negates all correlations in our detection statistic. We did not make any assumption on the distribution of the data or the model $z$ here. In Appendix D we give multiple alternate derivations of this result. Additionally, it follows straightforwardly from Eq. (D10). We have explored other special groups, such as the special orthogonal group $SO(M)$ instead of $U(M)$. In principle such other groups can also be used to scramble, and they destroy correlations similarly. But to not overcomplicate the discussion, we focus only on $U(M)$ in this paper.

### D. Phase scrambling

Phase scrambling [34] can be defined as,

$$z_a^{\mathrm{s}} = z_a^{\mathrm{obs}} e^{J\phi_a}, \tag{67}$$

where $z^{\mathrm{obs}}$ is the observed data, and $z^{\mathrm{s}}$ is the scrambled data. The scrambled data has the same amplitude $|z|$ but a random phase $\phi$ added. The phases $\phi_a$ are uniformly distributed: $\phi_a \sim \mathrm{Uniform}(0, 2\pi)$. The phase scrambling procedure can then be represented by a matrix $R_{ab} = \delta_{ab}e^{J\varphi_a}$, which we can insert into the detection statistic,

$$D(z^{\mathrm{s}}) = z^{\dagger}\mathbf{R}^{\dagger}\mathbf{Q}\mathbf{R}z. \tag{68}$$

It is straightforward to check that the scrambling operation can be applied to $\mathbf{W}$ in Eq. (47) as well, since the denominator is invariant under $\mathbf{R}$. Indeed, this group of phase scrambling operations is a subset of the group of unitary transformations defined in Sec. IV C. Even though we already proved that the detection statistic is zero in expectation under the entire unitary group, we verify explicitly that it is also true for this subset.

If we denote the average over all possible rotations as $\mathbb{E}_{\phi}[\cdot]$, we can evaluate,

$$\begin{aligned}\mathbb{E}_{\phi}[z^{\dagger}\mathbf{R}^{\dagger}\mathbf{Q}\mathbf{R}z] &= \sum_{ab}\mathbb{E}_{\phi}[Q_{ab}R^{*}_{aa}R_{bb}z^{*}_{a}z_{b}] \\ &= \sum_{ab}Q_{ab}|z_a||z_b|\mathbb{E}_{\phi}[\exp\left(J(\phi_a - \phi_b)\right)] \\ &= \sum_{ab}Q_{ab}|z_a||z_b|\delta_{ab} = 0.\end{aligned} \tag{69}$$

The average is an integral over $\phi_a, \phi_b \in [0, 2\pi]$, which is equal to zero unless $\phi_a = \phi_b$. In the last line we used that $\mathbf{Q}$ has zeros on the diagonal. The expectation value $\mathbb{E}[D(z^{\mathrm{s}})] = 0$, which is exactly what we had hoped to accomplish with the phase scrambling procedure.

### E. Sky scrambling

Sky scrambling [33,34] is similar to the setup of phase scrambling, but now the complex phase of the data is not scrambled. Instead, the pulsar position is changed. This makes it more difficult to track analytically what happens to the detection statistic: sky scrambling can be interpreted as a nonlinear transformation of the data that effectively just changes the optimal filter $\mathbf{Q}$ in such a way that $\mathbf{W}' = \mathbf{S}^{\dagger}\mathbf{W}\mathbf{S}$. While this approach leaves the numerical data (and thus $H_0$) invariant, it results in a nonlinear transformation of the quadratic filter $\mathbf{Q}$ since we also have to transform the denominator of Eq. (47).

Evaluating the expectation value of $D(z)$ under the scrambling operation has not been done yet in the literature. We investigate it in two steps. Firstly, we ignore the denominator of the quadratic filter in Sec. IV E 1. Then, we tackle the full sky scrambles in Sec. IV E 2.

#### *1. Sky scrambling average 1: No denominator*

Here we provide the intuition on which the original sky scrambling procedure was derived. We ignore the denominator of Eq. (47), and we consider only what happens to $\mathbf{W}$. We use $\hat{p}_a$ to denote the pulsar position unit vector on the two-sphere. Transformations that move $\hat{p}_a$ from one point of the two-sphere to another point on the two-sphere are part of the group called $SO(3)$: the special orthogonal group in three dimensions. The goal is now to evaluate what happens on average under all transformations in $SO(3)$. As before with the unitary scrambles, we have to assume the Haar measure when describing the density on $SO(3)$. Fortunately, for $SO(3)$ the interpretation is just a uniform distribution on the two-sphere.

We write $\mathbb{E}_{\mathrm{SO}(3)^M}$ for the ensemble average under the scrambling operation of sky scrambles of all $M$ pulsars. Even though we also use $\langle\cdot\rangle$ for ensemble averages in this manuscript, we thought that using $\mathbb{E}$ here allows us to keep track of which ensemble we are referring to in later sections with less clutter. We now observe that $W'_{ab}$, and by extension $\mu_{ab} = \mu(\gamma_{ab})$, only depends on the angle $\gamma_{ab}$ between pulsar $a$ and pulsar $b$. Under the scrambling operation, we know that $\cos\gamma_{ab} \sim \mathrm{Uniform}(-1, 1)$. Therefore,

$$\mathbb{E}_{\mathrm{SO}(3)^M}[(S(\mathbf{W}))_{ab}] = \int \mathrm{d}(\cos\gamma)\mu(\cos\gamma_{ab}), \tag{70}$$

where we write $S(\mathbf{W})$ for the transformed $\mathbf{W}$. Also, we note that for a Gaussian ensemble of GW sources isotropically distributed on the sky, we have,

$$\mu(\cos\gamma) = \sum_{l=0}^{\infty}(2l+1)C_l P_l(\cos\gamma) \tag{71}$$

where $P_l$ is a Legendre polynomial of order $l$, and the coefficients $C_l$ are [22],

$$C_l = \begin{cases} 0 & \text{if } l < 2, \\ \frac{(l-2)!}{(l+2)!} & \text{if } l \geq 2. \end{cases} \tag{72}$$

Because all $P_l(x)$ with $l > 0$ integrate to zero over the range $x \in [-1, 1]$, we immediately find that $S(\mathbf{W}) = 0$ on average over all sky scrambles (for an alternate derivation, see Appendix B). This shows that sky scrambling would indeed cancel correlations if it was a linear transformation of the data, just like we found with the random scrambling of the phase. However, it is not a linear transformation, and we need to make sure we include the transformation of the denominator of $\mathbf{Q}$. We do this in Sec. IV E 2.

#### *2. Sky scrambling average 2: Full quadratic filter*

Transforming the quadratic filter $\mathbf{Q}$ with a sky scramble will also change the denominator. This makes calculating the average a lot more difficult. We therefore start with a simple case: just two pulsars. When there are just two pulsars, then $W_{12} = W_{21} = \mu(\gamma_{12})$ is the only nonzero element of $\mathbf{W}$. Therefore, we only need to calculate the average of $Q_{12}$ under all sky scrambles in order to find the average detection statistic.

To start, we first note that the position of the first pulsar is a gauge freedom in our problem, as is the azimuthal angle of the second pulsar. The only free parameter that matters is $\cos\gamma_{12}$, which is distributed uniformly in the range $[-1, 1]$. This gives,

$$\begin{aligned}\mathbb{E}_{SO(3)^2}[Q_{12}] &= \mathbb{E}_{\cos\gamma_{12}}\left[\left(\frac{\mu(\gamma_{12})}{\mathrm{Tr}(\mathbf{N}^{-1}\mathbf{W}\mathbf{N}^{-1}\mathbf{W})^{1/2}}\right)\right] \\ &= \mathbb{E}_{\cos\gamma_{12}}\left[\frac{\mu(\gamma_{12})}{\sqrt{2}\mu(\gamma_{12})N_1^{-1/2}N_2^{-1/2}}\right] \\ &= \frac{\sqrt{N_1 N_2}}{\sqrt{2}},\end{aligned} \tag{73}$$

where $N_a$ is the diagonal noise term of the $a$th pulsar. We see here that for two pulsars, the detection statistic does not cancel under sky scrambling, unlike for phase scrambling or unitary scrambling. In general, the detection statistic does not vanish exactly under sky scrambling. We therefore investigate what happens when the number of pulsars increases, in the limit of $M \to \infty$.

Without loss of generality, we assume for the moment that we are interested in element $Q_{1M}$ of $\mathbf{Q}$, for an array of $M$ pulsars, with $a, b \in [1, M]$. As before, we are allowed to fix one of the pulsars to a particular position. In this case, we fix $\hat{p}_M$. We therefore only need to consider $\hat{p}_1$ when calculating $Q_{1M}$. The trace in the denominator can be written as,

$$\begin{aligned}\mathrm{Tr}(\mathbf{N}^{-1}\mathbf{W}\mathbf{N}^{-1}\mathbf{W}) &= \sum_{a\neq b} N_a^{-1}N_b^{-1}\mu^2(\hat{p}_a\cdot\hat{p}_b) \\ &= \frac{2\mu^2(\hat{p}_1\cdot\hat{p}_M)}{N_1 N_M} + 2\sum_{a=2}^{M-1}\frac{\mu^2(\hat{p}_1\cdot\hat{p}_a)}{N_1 N_a} \\ &\quad + \sum_{\substack{a,b=2\\ a\neq b}}^{M-1}\frac{\mu^2(\hat{p}_b\cdot\hat{p}_a)}{N_a N_b} \\ &= \kappa(\hat{p}_1) + \eta = \eta(1 + \kappa(\hat{p}_1)/\eta) \\ &= \eta(1+x)\end{aligned} \tag{74}$$

where we write $\mu^2(\hat{p}_a\cdot\hat{p}_b) = (\mu(\hat{p}_a\cdot\hat{p}_b))^2$ for clarity, and we defined,

$$\kappa(\hat{p}_1) = \frac{2\mu^2(\hat{p}_1\cdot\hat{p}_M)}{N_1 N_M} + 2\sum_{a=2}^{M-1}\frac{\mu^2(\hat{p}_1\cdot\hat{p}_a)}{N_1 N_a}, \tag{75}$$

$$\eta = \sum_{\substack{a,b=2 \\ a\neq b}}^{M-1} \frac{\mu^2(\hat{p}_b \cdot \hat{p}_a)}{N_a N_b}, \tag{76}$$

$$x = \frac{\kappa(\hat{p}_1)}{\eta}. \tag{77}$$

All the $\mu^2$ terms are of order $\mathcal{O}(1)$, and they are all positive. We further assume that all $N_a$ are roughly the same order of magnitude. This means that $x \sim \mathcal{O}(1/M)$, because there are $\mathcal{O}(M^2)$ terms in $\eta$, and only $\mathcal{O}(M)$ terms in $\kappa(\hat{p}_1)$. We can now expand $Q_{1M}$ as,

$$Q_{1M} \approx \frac{\mu(\hat{p}_1 \cdot \hat{p}_M)}{2N_1 N_M \sqrt{\eta}} \left(1 - \frac{1}{2}x + \mathcal{O}(x^2)\right). \tag{78}$$

We see here that as $M \to \infty$, $x \to 0$, and $Q_{1M}$ will be proportional to $\mu(\hat{p}_1 \cdot \hat{p}_M)$. As we saw in Sec. IV E 1, that term averages to zero under $SO(3)$. And the integral over $\hat{p}_a$ with $1 < a < M$ only affects terms in $\eta$ and terms that vanish with $M \to \infty$. With this, we have shown that sky scrambling cancels the detection statistic response in the limit of large number of pulsars $M$. In Appendix B we derive some more identities regarding sky scrambling that are relevant to this discussion.

### F. Match statistic

The idea of a single "scramble" is that the data or the model is modified in some way that negates the signal. When sky scrambling was first introduced, it was thought that the various scrambles need to be statistically independent in some way. The intuition is that scrambles that are too similar do not represent an independent contribution to the distribution of the background statistic. Indeed, it is the same data and the same detection statistic, so if the scramble is similar enough between two scrambles we will get a similar value for the detection statistic in those two scrambles. To quantify this, a "match statistic," $M$, was introduced,

$$M = \frac{\sum_{a\neq b} \mu_{ab}\mu'_{ab}}{(\sum_{a\neq b} \mu_{ab}\mu_{ab} \sum_{a\neq b} \mu'_{ab}\mu'_{ab})^{1/2}}, \tag{79}$$

where we write $\mu_{ab} = \mu(\gamma_{ab})$ for the original data, and $\mu'_{ab} = \mu(\gamma'_{ab})$ for the scrambled sky positions. The match statistic $M$ is an attempt to quantify the overlap between average correlations of the original sky positions and the new positions, with respect to the GWB correlations they would induce. The reasoning is as follows. On average, over many realizations of the Gaussian ensemble, $\mu_{ab}$ represents the correlation between pulsar $a$ and pulsar $b$. Then, with respect to the scrambled sky locations, the average correlations over many realizations of the Gaussian ensemble is $\mu'_{ab}$. The optimal detection statistic is the normalized inner product of the correlations between pulsars, so the match statistic $M$ was constructed to represent the *average* correlations between the data with pulsars at the original sky positions and the scrambled sky positions.

At first glance, this approach is intuitive. Indeed, the community has found guidance in $M$, and various papers set a threshold on $M$ for when a scramble is used. It is thought that scrambled positions that match too much with another scramble (e.g., $M \geq 0.1$) would not provide enough independent information [1,34]. Others pointed out that there are subtleties with the construction of $M$, and that it should be noise weighted [35].

In the construction of $M$, $\mu'_{ab}$ represents the average correlations between pulsars under the Gaussian ensemble. However, the scrambled sky positions do not come from a Gaussian ensemble: the scrambled sky positions come from uniform draws of pulsar positions $\hat{p}$ from $SO(3)^M$. The distribution of correlations under that group of transformations is therefore not Gaussian, and it is not clear that expected correlations like $\mu'_{ab}$ under that group behave similarly. Moreover, it is not clear how important it would be to have scrambles with some match statistic that is (close to) zero. In this work, instead of using the match statistic, we focus only on the condition of a vanishing average detection statistic under the scrambling operation.

## V. DISTRIBUTION OF THE SCRAMBLED STATISTIC

In Sec. IV we derived scrambling methods from first principles, and we checked that the average detection statistic response is zero under various scrambling techniques. This is enough to show that scrambling techniques do what they are supposed to do: negate the signal in the detection statistic. But it does not tell us whether the distribution can serve as a replacement of the distribution of the detection statistic under $H_0$.

The probability density function and the cumulative density function (CDF) of the detection statistic under scrambling have been well simulated in preparation for the 2023 PTA data releases of various projects ([4,13,38,39] and related publications). We know empirically that the detection statistic background distribution we get from scrambling does not equal the distribution under $H_0$ (we see this later in Fig. 2). This makes sense, they depend on a specific realization of data. In general, if data are generated under $H_0$, the scrambling background distribution is sometimes wider than the analytical background distribution under $H_0$, and sometimes it is narrower. In this section we analytically calculate the spread and the distribution of the detection statistic under the scrambling operation.

### A. Spread under $H_0$

If we assume $H_0$ to represent a model where $z \sim \mathcal{N}^{\mathbb{C}}(0, \mathbf{N})$, then we can directly calculate the spread of the detection statistic using Isserlis's theorem [40]. In Equation (A7) of Appendix A we find,[2]

$$\Delta^2_{H_0} \coloneqq \mathbb{E}_{H_0}[D(z,\theta)^2] = \mathrm{Tr}(\mathbf{QN})^2 + \mathrm{Tr}(\mathbf{QNQN}), \tag{80}$$

where $N_{ab} = \delta_{ab}(\hbar^2 + \sigma_a^2)$ under $H_0$. The spread of the distribution therefore simplifies to,

$$\begin{aligned}\Delta^2_{H_0} &= \mathbb{E}_{H_0}[D(z,\theta)^2] = \mathrm{Tr}(\mathbf{QNQN}) \\ &= \frac{\mathrm{Tr}(\mathbf{WN}^{-1}\mathbf{WN}^{-1})}{\mathrm{Tr}(\mathbf{WN}^{-1}\mathrm{WN}^{-1})} = 1,\end{aligned} \tag{81}$$

where we define $\Delta^2$ to represent the spread of the distribution. This gives us a benchmark to compare the scrambling spread with. Of course Eq. (81) was obvious, because we constructed the detection statistic to be unit variance.

### B. Phase scrambling spread

We now do the same calculation for phase scrambling [34]. We follow the same procedure as we used in Sec. IV D, where we define a phase scrambling matrix $S_{ab} = \delta_{ab} e^{J\varphi_a}$, so that we can write,

$$D(z^{\mathrm{s}}) = z^\dagger \mathbf{S}^\dagger \mathbf{QS} z. \tag{82}$$

This is a complex rotation in each observation (each frequency bin in real PTA analyses). Mathematically that group is called the unitary group of degree one $U(1)$, or the circle group. The total phase scrambling group can therefore be denoted as Cartesian product of $M$ copies of the circle group: $U(1)^M$, which is often referred to as the $M$ torus. The phase scrambling operator $S$ is an element of this group: $S \in U(1)^M$. For brevity, we denote $G = U(1)^M$.

The spread of the detection statistic under transformations in the $M$-torus is now,

$$\begin{aligned}\mathbb{E}_G[(z^\dagger \mathbf{S}^\dagger \mathbf{QS} z)^2] &= \sum_{abcd} Q_{ab} Q_{cd} \mathbb{E}_G[z_a^* z_b z_c^* z_d] \\ &= \sum_{abcd} Q_{ab} Q_{cd} r_a r_b r_c r_d \mathbb{E}_\phi \\ &\quad \times [\exp(J(\phi_b - \phi_a + \phi_c - \phi_d))],\end{aligned} \tag{83}$$

where we define the complex modulus $r_a = |z_a|$. It is important to note here that we are allowed to do this because the denominator of Eq. (47) is invariant under inclusion of the scrambling operator $S$. The expectation over complex phases is now just an integral $\frac{1}{2\pi}\int \mathrm{d}\phi$ over every phase. All those integrals vanish, unless the phases exactly cancel one another. This insight leads us to,

$$\begin{aligned}\Delta_G(z) &= \mathbb{E}_G[(z^\dagger \mathbf{S}^\dagger \mathbf{QS} z)^2] \\ &= \sum_{abcd} Q_{ab} Q_{cd} r_a r_b r_c r_d \\ &\quad \times (\delta_{ab}\delta_{cd} + \delta_{bc}\delta_{ad} - \delta_{ab}\delta_{bc}\delta_{cd})\end{aligned} \tag{84}$$

$$\Delta^2_{U(1)^M}(z) = \sum_{a,b} |z_a|^2 |z_b|^2 Q_{ab}^2, \tag{85}$$

where on the last line we have used the fact that $Q_{aa} = 0$.

### C. Unitary scrambling spread

In Appendix D we calculate the variance of the detection statistic under the Haar measure,

$$\Delta^2_{U(M)}(z) = \mathbb{E}_{U(M)}[D(z,\theta)^2] = \frac{|\tilde{z}|^4}{M(M+1)} \mathrm{Tr}(\tilde{\mathbf{Q}}^2) \tag{86}$$

where $\tilde{z}$ and $\tilde{\mathbf{W}}$ are the noise-weighted data and correlation matrix, as defined in Sec. IV A.

### D. Expectation of the spread

While $\Delta^2_{H_0}$ is the true spread of the detection statistic background distribution under $H_0$, the scrambling $\Delta^2$ quantities are evaluated with respect to a specific realization of data. If we assume that $z$ is described by $H_0$, we can calculate the average spread of those $\Delta^2$ quantities under $H_0$. Some of the above expressions of scrambling $\Delta^2$ depend on $|z|^4$. Using Appendix A, we find,

$$\mathbb{E}_{H_0}[|z|^4] = \mathbb{E}_{H_0}\left[\left(\sum\nolimits_{a,b} z_a^* z_a z_b^* z_b\right)^2\right] = \mathrm{Tr}(\mathbf{N}^2) + \mathrm{Tr}(\mathbf{N})^2. \tag{87}$$

However, $\tilde{\mathbf{N}}$ is the noise-weighted covariance, which is $\mathbf{I}$, so this simplifies to $M(M+1)$. Now we can use this in Equation. (86), and also fill in the expectation in Eq. (84). These give, combined,

$$\mathbb{E}_{H_0}[\Delta^2_{U(1)^M}] = 1, \tag{88}$$

$$\mathbb{E}_{H_0}[\Delta^2_{U(M)}] = 1. \tag{89}$$

Interestingly, we see that the spread under phase scrambling and unitary scrambling is identical to the null-hypothesis background of Eq. (81). While the variance under scrambling alone differs between scrambling procedures, when also incorporating the variation in the complex amplitude, we arrive at the exact same spread of the detection statistic.

### E. Unitary scrambling: Full distribution and *p*-values

For the generalized $\chi^2$ distribution of the detection statistic of Eq. (12) under $H_0$ we have a numerical approximation method called Imhof [31] to calculate $p$-alues, as suggested by Hazboun *et al.* [10]. Under the unitary scrambling operation, we can do something similar. As we show in Appendix D 4 b, we may write the detection statistic as,

$$D(z,\theta) = \tilde{z}^\dagger \tilde{\mathbf{Q}}\, \tilde{z} \tag{90}$$

$$D(z,\theta) = |\tilde{z}|^2 \sum_i w_i \lambda_i, \tag{91}$$

where in the last step we have used Eq. (D41). The elements of $\tilde{z}$ are distributed as $\tilde{z}_a \sim \mathcal{N}^{\mathbb{C}}(0,1)$. There, $w$ is a uniform Dirichlet distributed random variable under the scrambling operation,

$$w \sim \mathrm{Dirichlet}(1,1,\ldots,1), \tag{92}$$

and $\lambda_i$ are the eigenvalues of $\tilde{\mathbf{Q}}$. This means that, under Haar-distributed unitary scrambling operations, the detection statistic is a weighted uniform Dirichlet distribution. We have confirmed this fact numerically. It seems likely that $p$-values under this distribution can be calculated analytically, just like with the generalized $\chi^2$ of the optimal statistic under a Gaussian $H_0$. We have not yet been able to find an accurate method to do so. However, since the samples of a uniform Dirichlet distribution can be very efficiently drawn numerically, we can calculate accurate "semianalytical" $p$-values by just drawing a great number of samples using Eq. (91). Using an online method (do not store everything in memory, update results as you go) we were able to draw a sufficient number samples for any realistic significance level in under a minute for a 67-pulsar array.

[2]Note: when using real-values variables, Isserlis's theorem picks up an extra factor of 2 in these equations.

Now that we have an analytic description of the distribution of the detection statistic under unitary scrambling, we can also use the properties of the uniform Dirichlet distribution to derive the mean and spread. We do this in Appendix D 4 c. Using those expressions, we recover the results of Secs. IV C and V C.

## VI. DATA IN POLAR FORM

The scrambling operations we define in earlier sections have one key element in common: the scrambling operations $S(z)$ are transformations of the data that leave $H_0$ invariant, meaning that the complex norm $|z|$ is invariant under the transformations. Because of this, it is instructive to take a look at the data and the likelihood function in polar form,

$$p(z|\theta,H_0)\mathrm{d}z = \frac{1}{\det(2\pi\mathbf{N})} \exp\left(-\frac{1}{2} z^\dagger \mathbf{N}^{-1} z\right) \mathrm{d}z \tag{93}$$

$$= \prod_a \frac{1}{2\pi\lambda_a^2} \exp\left(-\frac{1}{2}\frac{|z_a|^2}{\lambda_a^2}\right) \mathrm{d}z_a, \tag{94}$$

$$p(r,\phi|\theta,H_0)\mathrm{d}r\mathrm{d}\phi = \prod_a \frac{r_a}{2\pi\lambda_a^2} \exp\left(-\frac{1}{2}\frac{r_a^2}{\lambda_a^2}\right) \mathrm{d}r_a \mathrm{d}\phi_a, \tag{95}$$

where we use the fact that under $H_0$ the covariance matrix $\mathbf{N}$ is diagonal with elements $N_{ab} = \delta_{ab}\sigma_a^2$ on the second line, we defined $\lambda_a = 2\sigma_a$, and we incorporated the Jacobian of the transformation with $\mathrm{d}z = r\mathrm{d}r\mathrm{d}\phi$. This follows straight from the definition $z_a = r_a \exp(J\phi_a)$.

We see from Eq. (94) that the distribution of data under $H_0$ does not depend on the complex phase. That is in agreement with our intuition that the data is uncorrelated under $H_0$. In Eq. (68) we saw that random phase scrambling gets rid of the correlations on average, meaning that on average under distribution of Equation (94) the detection statistic will not have a response, even if the data were generated under $H_S$.

### A. Polar coordinate *p*-values

Because Eq. (94) does not depend on the coordinate $\phi$, it is fully separable. Leaving out the measure, we have

$$p(r,\phi|\theta,H_0) = f(r|\theta)g(\phi) = \prod_a f_a(r_a|\theta) g_a(\phi_a), \tag{96}$$

$$f_a(r_a|\theta) = \frac{r_a}{\lambda_a^2} \exp\left(-\frac{1}{2}\frac{r_a^2}{\lambda_a^2}\right), \tag{97}$$

$$g_a(\phi_a) = \frac{1}{2\pi}. \tag{98}$$

The distribution $g_a(\phi)$ is constant over the entire domain $[0,2\pi]$. The distribution $f_a(r)$ is called the Rayleigh distribution, and it represents the variability of the complex amplitude under $H_0$.

The attractive feature of data in polar form is that we may argue in general about other forms of $H_0$, rather than the usual multivariate Gaussian distribution that is common in the PTA literature. For instance, even if our noise models are misspecified, or there are other ways in which our $H_0$ is incorrect, we typically assume that the data are statistically not correlated between pulsars. This means we may still assume that $g_a(\phi_a) = 1/(2\pi)$, even if $f_a(r_a)$ is no longer the Rayleigh distribution.

The $p$-value can now be calculated as the following integral over $r$ and $\phi$

$$P(D > D(z_{\mathrm{obs}},\theta)|H_0) = \int_{D(z,\theta) > D(z_{\mathrm{obs}},\theta)} \mathrm{d}r\mathrm{d}\phi\, p(r,\phi|\theta,H_0), \tag{99}$$

$$P(D > D(z_{\mathrm{obs}},\theta)|H_0) = \int \mathrm{d}r f(r|\theta) \int_{D(r,\phi,\theta) > D(z_{\mathrm{obs}},\theta)} \mathrm{d}\phi\, g(\phi). \tag{100}$$

### B. Interpreting scrambling with fixed model parameters

The decomposition in Eq. (99) gives us a deep relationship with scrambling operations. Scrambling is based on the assumption that the null hypothesis $H_0$ does not contain correlations between pulsars in the data. This means that the likelihood does not depend on the complex phase. That is represented by $g(\phi)$, which is a constant. The integral over $d\phi$ in Eq. (99) is an integral over all possible correlations. Looking back at Eq. (69), that is exactly the integral we take when we calculate the average over phase scrambling. Phase scrambling is evaluating only part of the integral we need to evaluate when calculating a $p$-value: the integral over $\phi$, not $r$. Unitary scrambling is an integral over all uncorrelated *and correlated* $\phi$, rather than only the factorized scrambles in $g(\phi)$. We found that the result is the same for the detection statistic.

Scrambling is different from using $H_0$, because we fix the realization of data, which means we effectively change $f(r)$. Instead of a general distribution that makes sense with respect to the noise model or other assumptions, we exchange $f(r)$ for a Dirac delta function centered around the values set in the data. This is an extremely unphysical model, and in light of Eq. (99) one may wonder what good such a choice would do. Surely we need to choose $f(r)$ as realistic as possible in order to get an accurate estimate of the $p$-value.

Fortunately, we have shown that the resulting detection statistic distributions are representative *on average*. We have shown that explicitly in Sec. V. Reflecting on Eq. (99) this becomes somewhat obvious: we have used the physical $g(\phi)$ when scrambling, so naturally if we take a weighted average over the physical $f(r)$ we would get the original background distribution back. Since PTAs typically have a large number of pulsars with multiple frequencies, we find it not surprising that the difference between the two distributions has not been considered alarming by the community. But we stress that scrambling does not result in a reliable background distribution.

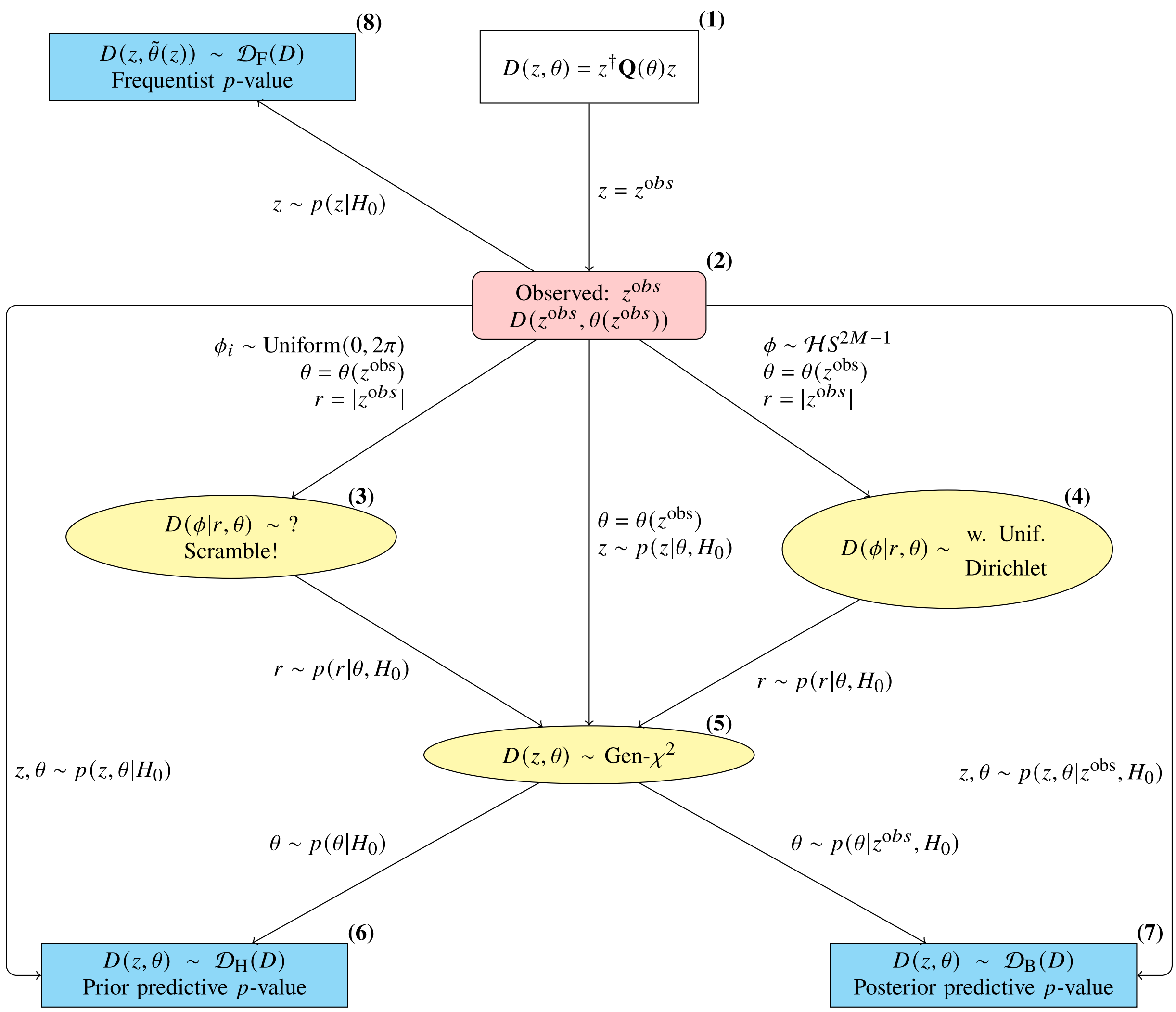

FIG. 1. Overview of how the distributions relate to one another. Red rounded nodes denote observations, white rectangular nodes denote definitions, yellow ellipses denote distributions, and the blue square nodes denote the background distribution of the detection statistic from which we calculate $p$-values. The numbers in parentheses correspond to the following nodes: (1) Detection statistic definition, (2) observed data, (3) phase shifting, (4) unitary scrambling, (5) generalized chi-squared distribution, (6) prior predictive distribution and $p$-value, (7) posterior predictive distribution and $p$-value, (8) frequentist distribution and $p$-value. Next to all the connectors the model/data assumptions are placed, which keep flowing through the diagram unless changed. So it can be seen here that the detection statistic under phase shifting in (3) changes into the General-$\chi^2$ of (5) if we change $r = |z^{\mathrm{obs}}|$ to $r \sim p(r, \phi, \theta, H_0)$.

### C. Drawing $r$ from $H_0$: Generalized $\chi^2$

Scrambling is a data transformation under which $H_0$ is invariant, which means that the complex amplitudes remain the same. If we allow the complex amplitude to be distributed according to $H_0$, we end up with a data distribution conditioned on model parameters $\theta$ and $H_0$: $p(z|\theta, H_0)$. The data is then distributed according to the likelihood. As discussed in Sec. III E, this makes the detection statistic follow a generalized $\chi^2$ distribution.

If we combine uniform phase scrambling of $\phi_i \sim \mathrm{Uniform}(0, 2\pi)$ and complex amplitude draws according to $r \sim p(r|\phi, \theta, H_0)$, it is trivial that we would end up with the same generalized $\chi^2$ distribution as when we would directly sample $z$ from the likelihood, because in polar coordinates $\phi$ and $r$ are independent, and $H_0$ is invariant under changes in $\phi$. It is less obvious that unitary scrambling combined with complex amplitude draws according to $r \sim p(r|\phi, \theta, H_0)$ yields the same generalized $\chi^2$ distribution. Indeed, the intermediate distributions of box (3) and box (4) in Fig. 1 are different. Interestingly, it is still true that we end up with the same generalized $\chi^2$ distribution.

The distribution of the detection statistic under unitary scrambling is a weighted uniform Dirichlet

$$D(z, \theta) = |\tilde{z}|^2 \sum_i \lambda_i w_i, \tag{101}$$

$$w \sim \mathrm{Dirichlet}(1, 1, \ldots, 1). \tag{102}$$

Here $|\tilde{z}|^2$ is the noise-weighted squared complex modulus of the data. Let $u = |\tilde{z}|^2$ be the squared complex modulus. Under $H_0$, $u$ is distributed as a $\chi^2(2M)$ distribution, which can be written as a gamma distribution with shape parameter $\alpha_0 = M$ and scale parameter $\theta = 2$. This allows us to define $u_i$,

$$u \sim \mathrm{Gamma}(M, 2), \tag{103}$$

$$u_i \coloneqq u w_i. \tag{104}$$

As we show in Appendix C 3 a, this means that $u_i$ is distributed as,

$$u_i \sim \mathrm{Gamma}(1, 2), \tag{105}$$

$$D(z, \theta) = \sum_i \lambda_i u_i. \tag{106}$$

This is a general property of the gamma distribution and the Dirichlet distribution. Note that gamma$(1, 2)$ is a $\chi^2$ distribution with two degrees of freedom, which means that the detection statistic is just a generalized $\chi^2$ distribution with weights $\lambda_i$. This is the same distribution we obtained by sampling from $p(z|\theta, H_0)$ directly, because the $\lambda_i$ are the same eigenvalues. It is very interesting to see that we obtain the same distribution, even though the intermediate distributions are slightly different. This corresponds to path $(2 \rightarrow 3 \rightarrow 5)$ and $(2 \rightarrow 4 \rightarrow 5)$ in Fig. 1, which we describe in Sec. VIII.

## VII. UNCERTAIN MODEL PARAMETERS

In all our discussions up to now, we have kept the model parameters $\theta$ fixed, or we assumed them to be known. Importantly, the detection statistic $D(z, \theta)$ depends on the model parameters because the noise covariance $\mathbf{N}$ depends on the model parameters. As we have shown in previous sections, scrambling operations can be analytically derived, and they depend on the same modeling assumptions and parameters. This subtle issue is not dealt with consistently in the PTA literature [1,2,36,41]. Moreover, some of the published results even keep certain model parameters fixed (set to zero) in their Bayesian analysis, which is an example of circular analysis [42]. In this section we discuss these subtleties in detail.

### A. Drawing from $H_0$

The main question that we need to address when constructing a $p$-value for GWB detection PTAs is "How does one sample data from $H_0$ for use in the detection statistic, without assuming too much about $H_0$?" There are various ways in which this can be done, including,

(i) Observe lots of noise-only data
(ii) Use simulations
(iii) Create a representative generative statistical model

Observing noise-only data is a gold standard if this can be done reliably. For instance, the time-slides used in LIGO [43] are a good example where such an approach is taken. Simulations can be used in a case where the data generating process is well understood, but too complex to model statistically. Chaotic systems of any sort fall into this category. A representative generative statistical model can be seen as the last resort, because such a statistical model necessarily involves many assumptions regarding the data and the noise model.

Scrambling methods are a way to replace observations of noise-only data. But, as we have shown in Sec. VI B, it is more correct to view scrambling methods as a more restrictive form of using a modeled $H_0$, where we fix the complex amplitudes $r$, and only integrate over the phase $\phi$ when calculating $p$-values. In this section we discuss what we should be doing instead of scrambling operations.

### B. Parameter variability

Most/realistic models for PTA data have some notion of noise parameters that can vary from pulsar to pulsar and from realization to realization: we do not know exactly what noise process governs the rotational stability of a pulsar, and there are many effects that are potentially introduced during the observational process. The somewhat philosophical aspect of this is whether the noise parameters need to be assumed fixed, or whether they can vary from realization to realization.

Even though Bayesians and Frequentists interpret variance in model parameters differently, both views can incorporate model parameters that vary from realization to realization. In the Frequentist interpretation, model parameters can be seen as being part of the data generating process,

$$\theta \sim p_f(\theta|H_0), \tag{107}$$

$$z \sim p(z|\theta, H_0). \tag{108}$$

Here we see that in order to generate data, we first need to draw a set of model parameters $\theta$, after which we can sample $z$. It is important to stress here that $p_f(\theta|H_0)$ should *not* be interpreted as a prior distribution. Rather, it is the distribution that represents variability under repeated experiments, which is completely in line with the Frequentist interpretation *if* we were able to do repeated experiments and those parameters indeed vary from experiment to experiment. In this view, $\theta$ should be seen as unobserved quantities in the data generation process, which can vary from realization to realization. However, this view of "$\theta$ has a latent unobserved part of the data generating process" breaks down in the Frequentist point of view if we need to *estimate* $\theta$ in order to do inference, which is the situation for PTA analysis. At that point $\theta$ is a model parameter that is assumed fixed.

Therefore, we adopt the Frequentist view that we simply leave out $\theta$ from the data generating process, and instead just draw data $z$ from $p(z|H_0)$. Mathematically this is equivalent to integrating the likelihood times sampling distribution $p(z|\theta, H_0)p(\theta|H_0)$ over $\theta$ if there is variability in model parameters, or keeping them fixed, which implicitly means we condition on them. In either case, we just write $z \sim p(z|H_0)$.

In practice, Frequentists can employ a variety of techniques to construct $H_0$ in the presence of uncertain $\theta$. Although we advocate Bayesian methods on the grounds of internal consistency, a much-used Frequentist method would be to "plug in" the estimate of the model parameters $\hat{\theta}(z^{\mathrm{obs}})$ as the parameter values for $H_0$, and generate data from $p(z|\hat{\theta}, H_0)$, while reestimating $\theta$ for each realization of data. This reestimation step is important, and is not routinely done in the PTA literature: it makes the Frequentist $p$-value computationally expensive.

The Bayesians view the model parameters subjectively, where the prior distribution $p(\theta|H_0)$ represents the belief in particular values of $\theta$. This can be summarized by using the joint prior predictive, which can also be interpreted as a generative model,

$$p(z, \theta|H_0) = p(z|\theta, H_0)p(\theta|H_0). \tag{109}$$

### C. Frequentist $p$-values

If we allow for uncertainty in model parameters under the generative model of data defined in Sec. VII B, we need to draw $z \sim p(z|H_0)$, whatever we define $H_0$ to be, and acknowledge that $\theta$ can have a different *estimated* value in each realization of $z$. The $p$-value is then calculated as,

$$P(D > D(z_{\mathrm{obs}})|H_0) = \int_{D(z,\tilde{\theta}) > D(z_{\mathrm{obs}},\hat{\theta})} \mathrm{d}z\, p(z|H_0). \tag{110}$$

Here we have used $\hat{\theta}$ as the estimated $\theta$ from the observed data, and $\tilde{\theta}$ as the estimated $\theta$ based on the data drawn from $H_0$ inside the integral. The use of $\hat{\theta}$ means that an optimization of the model parameters given the data $z_{\mathrm{obs}}$ has to be carried out for each realization of data. That makes the Frequentist statistic not a computationally cheap statistic.

It is very interesting to see what happens to the estimated $\tilde{\theta}$ from the realizations of $z$ under scrambling operations. Since $H_0$ is invariant under scrambling operations, and $\theta$ parametrizes the data under $H_0$, this means that all $\tilde{\theta}$ remain the same under scrambling operations. Therefore, scrambling operations cause the estimated model parameters to not vary under $H_0$. This is an important limitation of scrambling operations that can have significant consequences on the $p$-value.

As an example, take the MeerKAT pulsar timing array second data release [44], where the authors use a "codified analysis" to select which noise contributions to turn off (i.e., fix, set to zero, and not vary) during their final GWB analysis [36]. This model with fixed noise parameters is called the "Data Model." There is also an identical model where all noise parameters are varied, called the "ER Model" (for extended red noise). The only difference is that certain noise parameters are not varied during the final Bayesian analysis. They found a profound difference in their detection statistic, and the authors chose only to use the data model with certain parameters fixed. The authors argued that this was well motivated, because their scrambling analysis also showed a significant detection statistic. Moreover, we agree with the authors that the cross-correlation plot looks *very* compelling. However, as we show here, scrambling operations inherently fix the model parameters, so it is only natural that a model with fixed model parameters would yield the same seemingly significant result as an analysis based on scrambling methods.

We therefore conclude that, because we need to incorporate the uncertainty in the model parameters $\theta$, we need to draw $z \sim p(z|H_0)$ and reestimate $\tilde{\theta}$ for each realization. Scrambling operations are incompatible with this view by construction, and their use has likely been overstated in the literature. So far, no Frequentist $p$-values have been reported in the literature that incorporate uncertainty in model parameters.

### D. Prior predictive $p$-values

Following a similar line of thought as when we discussed the Frequentist $p$-value above, we can interpret the uncertain model parameters from a Bayesian perspective. Basically that means we interpret the sampling distribution of Eq. (107) as the *prior*. Taking that approach, the data and model parameters are sampled from the joint prior predictive distribution [45], $(z,\theta) \sim p(z,\theta|H_0)$. From that we can find the prior predictive $p$-value,

$$p(z,\theta|H_0) = p(z|\theta,H_0)p(\theta|H_0), \tag{111}$$

$$P(D > D(z_{\mathrm{obs}})|H_0) = \int_{D(z,\theta)>D(z_{\mathrm{obs}},\theta)} \mathrm{d}z\mathrm{d}\theta\, p(z,\theta|H_0). \tag{112}$$

While the prior predictive $p$-value is Bayesian in nature, it does not take into account the fact that the observations would update our knowledge about the model parameters. Therefore, the prior predictive is taken under a distribution that is wider than our current knowledge would suggest. In our testing, we found the prior predictive distribution in combination with our detection statistic result in a $p$-value that does not allow us to discriminate $H_0$ from $H_S$, which makes it of very limited use.

### E. Posterior predictive (Bayesian) $p$-values

The $p$-values in a Bayesian approach are calculated by acknowledging the implications the data has on our knowledge of the model parameters by updating our prior beliefs with the observations. Therefore, the Bayesians use the joint posterior predictive when calculating the $p$-value,

$$p(z,\theta|z^{\mathrm{obs}},H_0) = p(z|\theta,H_0)p(\theta|z^{\mathrm{obs}},H_0), \tag{113}$$

$$P(D > D(z_{\mathrm{obs}})|z^{\mathrm{obs}},H_0) = \int_{D(z,\theta)>D(z_{\mathrm{obs}},\theta)} \mathrm{d}z\mathrm{d}\theta\, p(z,\theta|z^{\mathrm{obs}},H_0). \tag{114}$$

We see here that the joint posterior predictive is the product of the likelihood and the posterior. The procedure of Eq. (113) was very clearly described by [PP1 [11] ], where the authors also make the connection with the noise-marginalized optimal statistic (NMOS [27]). The NMOS is carrying similar integrals as in Eq. (113), for a posterior predictive $p$-value the $p$-value is the integrand, not the detection statistic (referred to as $S/N$ in PP1). In [46] the PP1 Bayesian $p$-value is calculated on the NANOGrav 15 yr data, which came out to be an equivalent of a Gaussian $3.2\sigma$ significance. That $p$-value is the only statistically rigorous $p$-value in the PTA literature to date. Note that PP1 did not have a way to *directly* calculate the integral of Eq. (113) from the data because the authors did not have a way to use the analytical generalized $\chi^2$ distribution. We show in Sec. IX how to efficiently calculate the posterior predictive $p$-value directly from the data using a standard Bayesian Markov chain Monte Carlo (MCMC) analysis.

## VIII. RELATIONSHIP BETWEEN ALL DISTRIBUTIONS

In this manuscript we discuss how the detection statistic behaves under various transformations and assumptions regarding the distribution of the input parameters. It is instructive to visualize how all these assumptions fit in the larger picture, which we have illustrated in Fig. 1. In this section we describe all distributions, and discuss how these distributions are related.

### A. The observed detection statistic

In Fig. 1 we see that we get to the observed detection statistic (number 2) from the detection statistic (number 1) only by following the arrow next to which it says that we are fixing/setting the data $z = z^{\mathrm{obs}}$. This is a slight simplification: the detection statistic depends also on the model parameters $\theta$. Therefore, in box number (2), we have written $\theta(z^{\mathrm{obs}})$ for the model parameters. However, we note that this is the Frequentist way of using the data and the model parameters in the detection statistic. Bayesians would not take this approach.

### B. The Frequentist $p$-value

A true Frequentist $p$-value is derived by drawing realizations $z \sim p(z|H_0)$ and determining how often we encounter a detection statistic larger than what we found on the real data. Since our detection statistic depends on the model parameters, it is important that we reestimate the model parameters from the realizations of data: $\tilde{\theta} = \tilde{\theta}(z)$.

This gives box number (8). We can think of this procedure as being part of the detection statistic, which in principle changes the detection statistic from a quadratic estimator of the data into something that is not a quadratic estimator of the data. Estimating the model parameters from the data is typically done with a maximum likelihood estimate or some other point estimator. We have denoted the distribution of the detection statistic in the Frequentist approach as $\mathcal{D}_{\mathrm{F}}(D)$.

### C. Phase scrambling and unitary scrambling

When we take the observed data, then fix the model parameters to $\hat{\theta} = \theta(z^{\mathrm{obs}})$, and we randomly adjust the complex phases of the observed data by drawing them from some distribution, we are effectively carrying out a scrambling operation. This is what is shown with the arrows from box number (2) in Fig. 1 to box numbers (3) and (4). Drawing the phase parameters $\phi_i$ independently from a uniform distribution, we carry out the phase shifting procedure introduced by Taylor *et al.* [34], which is shown by box number (3). If we instead draw the phases uniformly from a complex $M$-sphere assuming a Haar measure, we do the equivalent of unitary scrambling as derived in Sec. IV shown in box (4).

We showed in Sec. V E that the detection statistic becomes distributed as a weighted uniform Dirichlet distribution under unitary phase scrambling using Haar-distributed unitary transformations on the noise-weighted data, provided we keep the model parameters and the complex amplitudes of the data fixed. The detection statistic becomes similarly distributed under uniform random phase scrambling (phase shifting). However, we have not been able to analytically identify the distribution of the detection statistic under uniform phase scrambling. We know it is a different distribution from the weighted uniform Dirichlet, because the variance of the background distributions of Eqs. (84) and (86) are different. Visually, the distributions seem indistinguishable, as can be seen in Fig. 2.

### D. The prior/posterior predictive

The full distribution of the detection statistic under the prior predictive and posterior predictive also needs to include variability of the model parameters. The precise value of the noise parameters are unknown, and with repeated experiments under $H_0$ these noise parameters will vary. We found the prior predictive to be of little value for our detection statistic. Bayesians will acknowledge that the observations $z^{\mathrm{obs}}$ update our prior beliefs on $\theta$, and instead draw theta from the posterior: $\theta \sim p(\theta | z^{\mathrm{obs}}, H_0)$. Starting from the generalized $\chi^2$ of box (5), this gives us box (6) for the prior predictive and box (7) for the posterior predictive in Fig. 1. Alternatively, we obtain the detection statistic distribution immediately by drawing from either the joint prior predictive $z, \theta \sim p(z, \theta | H_0)$ or the joint posterior predictive $z, \theta \sim p(z, \theta | z^{\mathrm{obs}}, H_0)$ (Bayesians). We have denoted the prior predictive detection statistic distribution as $\mathcal{D}_{\mathrm{H}}(D)$ and the posterior predictive detection statistic distribution as $\mathcal{D}_{\mathrm{B}}(D)$.

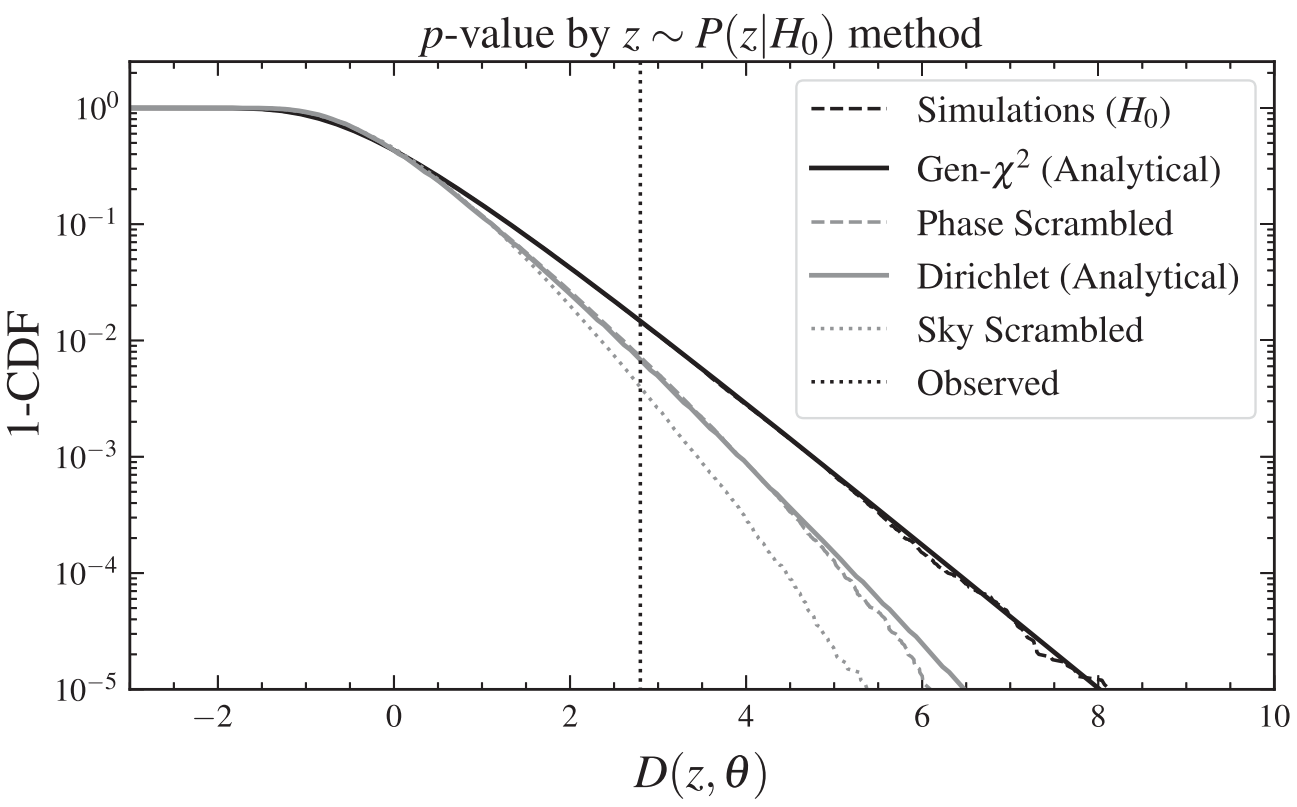

FIG. 2. Visual representation of the CDF of the distribution of the detection statistic for a 67-detector configuration, with pulsars positioned in the locations of the NANOGrav 15-year pulsars. The model is as described in Sec. III A, with $\sigma_a^2 \sim \mathrm{UNIFORM}(\frac{1}{2}, 1)$ and $\hbar^2 = 1$. The model parameters are known. The simulations (black dashed) were $z \sim p(z|H_0)$, sky scrambles (gray dotted) and phase scrambles (gray dashed) are done as described in the main text. We see that the generalized $\chi^2$ distribution (black solid) follows the simulations exactly. We also see that phase scrambling follows the analytical Dirichlet (gray solid) quite well, up until numerical deviations seem to take hold. For an observed detection statistic (black dotted), only the generalized $\chi^2$ and the simulations from $H_0$ would result in a correctly estimated $p$-value.

### E. Simple example

As an example of what the various distributions in Fig. 1 look like, we create a simulated experiment with $H_0$ and $H_S$ based on the model of Sec. III A. We choose a relatively strong signal scenario with $\sigma_a^2 \sim \mathrm{UNIFORM}(\frac{1}{2}, 1)$ and $\hbar^2 = 1$, because we only have one complex observation per pulsar for simplicity. We assume the noise parameters are fixed and known, because that allows us to really showcase what the differences are between scrambling operations and the true $H_0$ and distributions.[3]

For this experiment, we visualize the results in Fig. 2, where we consider various ways to draw $z \sim p(z|H_0)$ or to create the detection statistic distribution: simulations by drawing $z$ from the true $H_0$ (blue solid), phase scrambling (orange solid), sky scrambling (purple solid), analytical generalized $\chi^2$ distribution (gray dash-dotted), and semi-analytical weighted uniform Dirichlet (dashed). For reference, we also generate a single dataset from $H_S$, and indicate the observed detection statistic (vertical red dashed line). The simulations and scrambles all used one million realizations of data. We see that the simulations follow the generalized $\chi^2$ very accurately, and the phase scrambles follow the weighted uniform Dirichlet distribution very faithfully. At the higher range of $p$-values the realization-based methods start to deviate from the analytical results due to low-number statistics.

[3]We have done many types of experiments, with varying complexity, and in the end the simplest example is the most educational.

These results are exactly as expected: we see complete agreement with analytical results, and we also see a clear difference between the generalized $\chi^2$ and the weighted uniform Dirichlet. In this simple example there are no other unknowns: all parameters are known and there is no model misspecification. Any method to approximate the background distribution of the detection statistic should pass at least this test. Only the simulations and the generalized $\chi^2$ seem to yield truthful estimates of the $p$-value. In the literature similar tests have been carried out in the run-up to the various 2023 PTA data releases. We believe these differences between background estimation methods were not alarming due to the experimental setup: those experiments were set up to be realistic and many other effects were included in the simulations. These effects limited the resolution with which the differences between the distribution could be noticed. For instance, if we increase the number of observations (frequency bins) per pulsar, the differences *can* get smaller. Fitting for model parameters can partially mitigate the differences between the distributions. We stress that, although differences can get smaller in realistic settings, this is not guaranteed to be true, and indeed in Fig. 3 of Agazie *et al.* [1] we do see a difference between the simulations and the analytical background distribution.

## IX. CALCULATING $p$-VALUES: USING GENERALIZED $\chi^2$ DISTRIBUTIONS

Now that we have a solid understanding of the mechanics of $p$-values from the optimal detection statistic in PTAs, we are in a position to work out the details of how to calculate $p$-values in practice. In brief: we need to evaluate Eq. (113). As described in Sec. III E, the distribution of the detection statistic for fixed model parameters is a generalized $\chi^2$ distribution. Given the data, the $p$-value is then the expectation value of that $p$-value over the posterior distribution of the model parameters.

While the generalized $\chi^2$ distribution has been properly introduced in Hazboun *et al.* [10], their calculations were carried out in the time domain. With contemporary datasets that means the quadratic filter of the detection statistic has on the order of trillions of elements ($\sim 10^6 \times 10^6$), which makes the required eigen decomposition intractable. In this section we derive a practical method to calculate the generalized $\chi^2$ distribution for modern datasets in an efficient way. Then, using that, we show how to calculate a rigorous posterior predictive $p$-value for the PTA detection statistic. The same approach can be used to calculate a Frequentist $p$-value.

### A. The rank-reduced PTA model

While the full PTA model is typically described in terms of a likelihood function with many model components, we simplify the discussion in this work by focusing only on the covariance matrix. In the previous sections we have used a toy model where the signal and noise were described by zero-mean Gaussian random variables that were either correlated between pulsar ($H_S$) or uncorrelated between pulsar ($H_0$). Without loss of generality, we stick to zero-mean Gaussian random variables where we use so-called rank-reduced methods to express our covariance matrices [47,48]. This boils down to the following:

$$\mathbf{N} = \mathbf{W} + \mathbf{T}\mathbf{B}_0\mathbf{T}^T, \tag{115}$$

where $\mathbf{N}$ is the full time-domain ($n \times n$) covariance matrix under $H_0$, $\mathbf{T}$ is a $(n \times m)$ matrix of basis functions, where $n$ is the total number of observations and $m$ is the number of basis functions in $\mathbf{T}$. $\mathbf{W}$ is a full-rank covariance matrix that is quick to invert, usually representing "white noise." Typically $m \ll n$, which generates a large computational gain through the application of the Sherman-Morrison-Woodbury matrix inversion lemma [49,50]. For the $H_S$ hypothesis we use $\mathbf{B}_1$ instead of $\mathbf{B}_0$ in the above equation. For our generalized $\chi^2$ $p$-value derivation, we assume that $\mathbf{B}_{0,1}$ is the only difference between $H_0$ and $H_S$, where both are required to be symmetric and positive definite. The white noise matrix $\mathbf{W}$ is assumed to be the same and constant in both hypotheses. Varying white noise models are sometimes considered in the PTA literature, but for GWB searches they are typically held fixed for computational efficiency. This is a good approximation, because the parameters that describe $\mathbf{W}$ are not expected to be covariant with any GWB parameters.

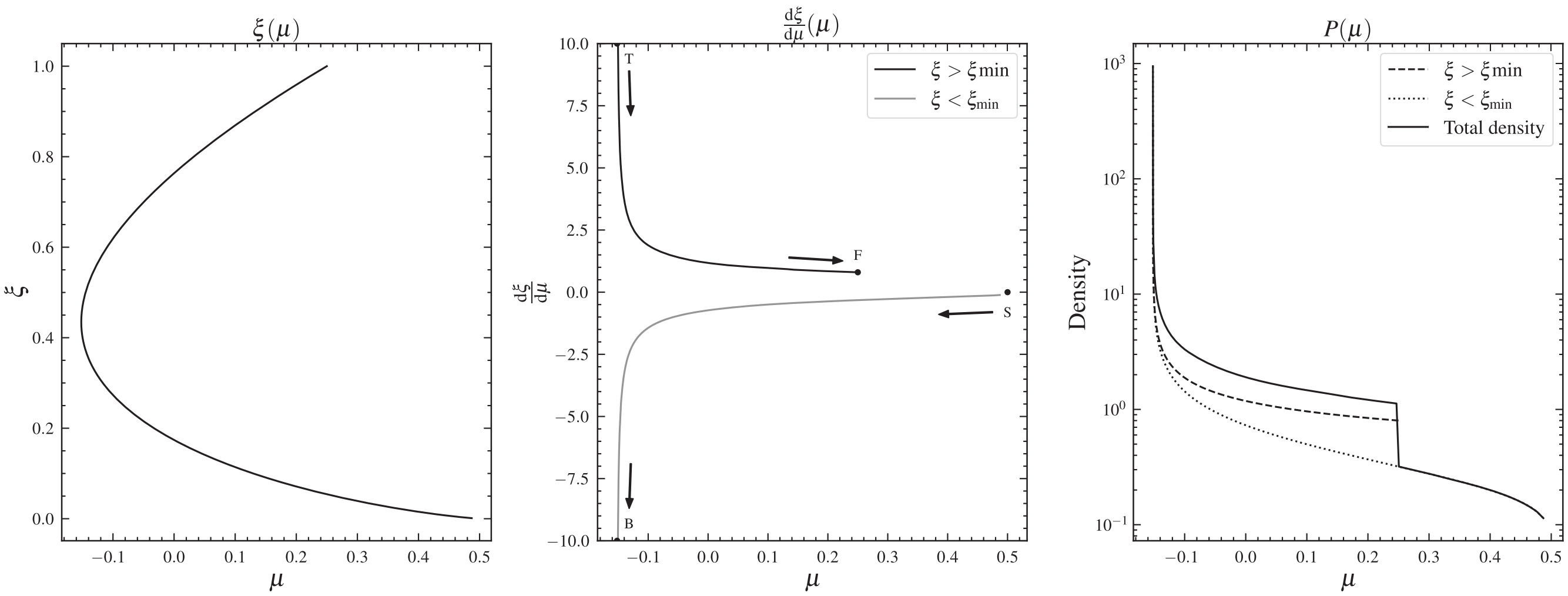

FIG. 3. Density of the H&D correlations under uniform sky scrambling. Left panel: the pulsar separation $\xi$ as a function of H&D correlations $\mu$. Middle panel: the derivative of the inverse correlations with the integration bounds $(T, B, S, F)$ marked. Integration path $S \rightarrow B$ represents the integral over the part of the H&D curve with small angular separations $\xi > \xi_{\text{min}}$. Integration path $T \rightarrow F$ represents the integral over the part of the H&D curve with large angular separations $\xi > \xi_{\text{min}}$. Right panel: the total H&D correlation density under uniform sky scrambles, presented as the sum of the two components.

### B. The white noise matrix W

First we describe the matrix $\mathbf{W}$. While various choices for $\mathbf{W}$ are used in different implementations (e.g., MarginalizingTimingModel in ENTERPRISE [51]), here we assume that $\mathbf{W}$ only contains the so-called white noise terms. This means that $\mathbf{W}$ is block diagonal, with each block corresponding to a single observation epoch for a single pulsar. The off-diagonal elements in a block correspond to the "ECORR" (see, e.g., [52]) modeling component. More concretely,

$$\mathbf{W} = \begin{pmatrix} \mathbf{W}_1 & 0 & \cdots & 0 \\ 0 & \mathbf{W}_2 & \cdots & 0 \\ \vdots & \vdots & \ddots & \vdots \\ 0 & 0 & \cdots & \mathbf{W}_{N_e} \end{pmatrix}, \tag{116}$$

with $N_e$ the number of epochs, and

$$W_{a,jk} = \sigma_{a,j}^2 \delta_{jk} + \mathcal{J}_a^2 \tag{117}$$

$$= \sigma_{a,j}^2 \delta_{jk} + \mathcal{J}_a^2 u_j u_k \tag{118}$$

where $\mathbf{W_a}$ is the $a$th block of $\mathbf{W}$, $\sigma_{a,j}$ is the standard deviation of the $j$th observation of the $a$th epoch, and $\mathcal{J}_a$ represents the ECORR term that is fully correlated among all observations of the $a$th epoch. In the second line we have expressed the ECORR term in terms of a rank-1 matrix, where $u_j$ is a vector of all ones $u = (1, 1, \ldots, 1)$. This allows us to express the inverse and Cholesky factor of $\mathbf{W}_a$ in terms of rank-one updates to a diagonal matrix. The inverse of each $\mathbf{W}_a$ is given by the Sherman-Morrison rank-one update formula,

$$W_{a,jk}^{-1} = \sigma_{a,j}^{-2}\delta_{jk} - \frac{\sigma_{a,j}^{-2}\sigma_{a,k}^{-2}}{(\sum_j \sigma_{a,j}^{-2} + \mathcal{J}^{-2})}. \tag{119}$$

The Cholesky factor of $\mathbf{W}_a$ can be similarly calculated with a block-wise rank-one update [53]. We have implemented the rank-one Cholesky update in the ENTERPRISE and FASTSHERMANMORRISON [54] packages, and we refer to that code for the details. The Cholesky rank-one update is implemented in terms of a "sqrtsolve" function, where the inverse of the Cholesky factor is multiplied with a column vector or a matrix. In principle, that solve function can be implemented for any matrix of the form of Eq. (115). However, numerically stable and efficient methods are not available in general. The result is that, for computational reasons, we require the ENTERPRISE PTA model to be built using the TimingModel modeling component instead of the often-used MarginalizingTimingModel component, because that keeps our $\mathbf{W}$ in the shape described above.

### C. The quadratic filter for a PTA

Written using the above notation, we can define the optimal quadratic filter that allows us to distinguish $H_0$ from $H_S$ given the model model parameters,

$$\mathbf{Q} = \frac{\mathbf{N}^{-1}\mathbf{T}\Delta\mathbf{B}\mathbf{T}^T\mathbf{N}^{-1}}{\sqrt{2\mathrm{Tr}[\mathbf{N}^{-1}\mathbf{T}\Delta\mathbf{B}\mathbf{T}^T\mathbf{N}^{-1}\mathbf{T}\Delta\mathbf{B}\mathbf{T}^T]}}, \tag{120}$$

where we write $\Delta\mathbf{B} = \mathbf{B}_1 - \mathbf{B}_0$. As before, beware of the factor of 2 in the square root in the denominator. This factor arises because we use real-valued variables in real PTA applications (we expand in sines and cosines), which causes Isserlis's theorem of Eq. (A7) to pick up an extra factor of two. The detection statistic is now,

$$D(y, \theta) = y^T\mathbf{Q}y, \tag{121}$$

where we denote the real-valued concatenation of all data as $y$.

As described in Sec. III E and Hazboun *et al.* [10], the distribution of the detection statistic under $H_0$ can be found by whitening the data using the matrix square root of the $H_0$ data covariance. Our starting point is the data as is, for which the data covariance under $H_0$ is given by,

$$\langle yy^T \rangle_{H_0} = \mathbf{N}_0 = \mathbf{W} + \mathbf{T}\mathbf{B}_0\mathbf{T}^T. \quad (122)$$

The data covariance matrix of Eq. (122) is a full-rank matrix. As described by Hazboun *et al.* [10], we can use the Cholesky decomposition or eigen decomposition to find the square root of the covariance matrix. Low-rank updating methods are not numerically stable enough to be used on a basis with the number of columns that will be present in practice in $\mathbf{T}$. This was also realized by Hazboun *et al.* [10], who used a brute-force eigen decomposition of the per-pulsar covariance matrix and the full quadratic filter. This is not feasible for modern datasets, which can have $n \sim 10^6$.

Our method to find the analytic distribution of $D(y, \theta)$ under $H_0$ is to introduce a chain of linear coordinate transformations that step-by-step whiten and compress Eq. (120) without loss of information.

#### 1. Transformation 1: Whitening the white noise

What is referred to as "white noise" in the PTA literature is actually a nonstationary random process with a block-diagonal covariance matrix structure given by $\mathbf{W}$. As described above, we have implemented an efficient Cholesky decomposition "solve" routine that effectively decomposes $\mathbf{W} = \mathbf{L}_\mathrm{W}\mathbf{L}_\mathrm{W}^T$ where $\mathbf{L}_\mathrm{W}$ is a lower-triangular matrix. We use this as our first transformation,

$$y_1 = \mathbf{L}_\mathrm{W}^{-1} y \quad (123)$$

this turns the data covariance into,

$$\mathbf{N}_1 = \langle y_1 y_1^T \rangle_{H_0} = \mathbf{I}_n + \mathbf{T}'\mathbf{B}_0\mathbf{T}'^T, \quad (124)$$

where we use the notation $\mathbf{T}' = \mathbf{L}_\mathrm{W}^{-1}\mathbf{T}$.

#### 2. Transformation 2: First compression of the data

The data $y_1$ and the covariance matrix $\mathbf{N}_1$ are still the same size as before. Next we will reduce the size of the data and covariance without changing the detection statistic. Let us first write the detection statistic in the basis of $y_1$,

$$D(y_1, \theta) = y_1^T(\mathbf{I}_n + \mathbf{T}'\mathbf{B}_0\mathbf{T}'^T)^{-1}\mathbf{T}'\Delta\mathbf{B}\mathbf{T}'^T(\mathbf{I}_n + \mathbf{T}'\mathbf{B}_0\mathbf{T}'^T)^{-1}y_1/\mathcal{n}, \quad (125)$$

where for clarity we substitute $\mathcal{n}$ for the normalization in the denominator of Eq. (120). We observe that the quadratic filter has a particular structure: both on the left and the right we "weight" the data with the inverse covariance matrix under $H_0$, and in between those we have a low-rank filter in the basis $\mathbf{T}'$. The low-rank filter in the middle has the same basis ($\mathbf{T}'$) as the "update" term in the inverse covariance matrices under $H_0$. Because we have whitened the white noise matrix, we are able to use some tricks involving projection matrices. First we construct a projection matrix $\mathbf{P}_T$ with the properties,

$$\mathbf{P}_T = \mathbf{P}_T^2, \quad (126)$$

$$\mathbf{P}_T = \mathbf{G}_T\mathbf{G}_T^T, \quad (127)$$

$$\mathbf{P}_T\mathbf{T}' = \mathbf{T}', \quad (128)$$

$$\mathbf{G}_T^T\mathbf{G}_T = \mathbf{I}_m \quad (129)$$

where $\mathbf{G}_T$ is an $(n \times m)$ matrix. So, $\mathbf{G}_T$ and $\mathbf{T}$ are rectangular matrices of the same size. The matrix $\mathbf{G}_T$ can be found using a "thin" singular value decomposition (SVD): $\mathbf{T}' = \mathbf{G}_T\Sigma\mathbf{V}^T$, where $\Sigma$ is an $(m \times m)$ matrix consisting only of the nonsingular values of $\mathbf{T}'$.

Next, we observe that Eq. (125) will remain the same if we insert the projection matrix $\mathbf{P}_T$ to the left of $\mathbf{T}'$ and to the right of $\mathbf{T}'^T$. We also note that $\mathbf{N}_1$ commutes with $\mathbf{P}_T$: $[\mathbf{P}_T, \mathbf{N}_1] = 0$. These two facts lead us to transformation 2: a linear compression of the data,

$$y_2 = \mathbf{G}_T^T y_1 = \mathbf{G}_T^T\mathbf{L}_\mathrm{W}^{-1} y, \quad (130)$$

$$\mathbf{N}_2 = \mathbf{G}_T^T\mathbf{N}_1\mathbf{G}_T = \mathbf{I}_m + \mathbf{G}_T^T\mathbf{T}'\mathbf{B}_0\mathbf{T}'^T\mathbf{G}_T. \quad (131)$$

The detection statistic is invariant under the above linear data compression transformation, but the dimensionality of the data $y_2$ and the covariance $\mathbf{N}_2$ have been greatly reduced.

#### 3. Transformation 3: Full whitening of the data

Now that the data is of a manageable size per pulsar, we can whiten with a proper matrix square root. Even though the matrix $\mathbf{N}_2$ is formally a symmetric positive definite matrix, in practice that matrix is numerically ill conditioned because the timing model parameters are usually given nonphysical improper priors [e.g., [52,55] ], which in practice means that some of the diagonal elements of $\mathbf{B}_0$ are set to $10^{40}$ [51,56] leading to large condition numbers. Therefore, we instead first use the Sherman-Morrison-Woodbury matrix inversion lemma on $\mathbf{N}_2$ and we subsequently use an SVD to find the (non)singular values and the eigen decomposition. The singular values also immediately give us access to an (inverse) matrix square root to whiten the data and the model,

$$\mathbf{L}_B\mathbf{L}_B^T = \mathbf{N}_2, \quad (132)$$

$$y_3 = \mathbf{L}_B^{-1} y_2 = \mathbf{L}_B^{-1}\mathbf{G}_T^T\mathbf{L}_\mathrm{W}^{-1} y, \quad (133)$$

$$\mathbf{N}_3 = \langle y_3 y_3^T \rangle_{H_0} = \mathbf{I}_m, \quad (134)$$

$$D(y, \theta) = y_3^T\mathbf{L}_B^{-1}\mathbf{G}_T^T\mathbf{L}_\mathrm{W}^{-1}\mathbf{T}\Delta\mathbf{B}\mathbf{T}^T\mathbf{L}_\mathrm{W}^{-T}\mathbf{G}_T\mathbf{L}_B^{-T} y_3/\mathcal{n}. \quad (135)$$

In a way we have reached our objective already: the transformed data covariance matrix is white, and we can use the full eigenvalue decomposition of the quadratic filter already to find the distribution of $D(y, \theta)$. However, the vectors and matrices would still be rather large for a full array of pulsars. With 100 pulsars and moderately complex modeling choices the matrix we would have to take an eigenvalue decomposition of would still be $(30,000 \times 30,000)$ in size. Not prohibitive, but it can take well over an hour on a modern workstation's central processing unit. We therefore need one more compression transformation.

#### *4. Transformation 4: Second compression of the data*

The quadratic filter for the data $y_3$ is,

$$\mathbf{Q}_3 = \mathbf{L}_B^{-1}\mathbf{G}_T^T\mathbf{L}_\mathrm{W}^{-1}\mathbf{T}\Delta\mathbf{B}\mathbf{T}^T\mathbf{L}_\mathrm{W}^{-T}\mathbf{G}_T\mathbf{L}_B^{-T}. \tag{136}$$

We note that this is typically a low-rank matrix, because the cross-correlations are only defined for the frequency modes where we model the GWB. There are no cross-correlations between pulsars for DM variations, the timing model, or higher frequency modes. We say "typically," because in principle our formalism can be used to do model selection with an optimal statistic for any two hypotheses, not just between a CURN model and a model that does include correlations. However, for a CURN vs GWB model, the matrix $\mathbf{Q}_3$ is low rank: we can write the filter in terms of the bases $\mathbf{F}$, where $\mathbf{F}$ consists only of the *subset* of columns of $\mathbf{T}$ where the GWB has correlations. This makes $\mathbf{F}$ an $(n \times k)$ matrix, with $k < m < n$. If $b \in [1, m]$ are the column indices for $\mathbf{T}$, and $c \in [1, k]$ are the column indices of $\mathbf{F}$, then we can define the corresponding indices of $\mathbf{F}$ in $\mathbf{T}$ as $b = b(c)$. The $c$th column vector of $\mathbf{F}$ is then the $b(c)$th column vector of $\mathbf{T}$.

We wish to exploit the low-rank property of $\mathbf{Q}_3$ and introduce yet another set of projection matrices that can reduce our dataset even further. Similar to what we did before, we insert projection matrices into the quadratic filter. The projection matrix we introduce is $\mathbf{P}_F = \mathbf{G}_F\mathbf{G}_F^T$, which has the following properties:

$$\mathbf{P}_F = \mathbf{P}_F^2, \tag{137}$$

$$\mathbf{P}_F = \mathbf{G}_F\mathbf{G}_F^T, \tag{138}$$

$$\mathbf{T}\mathbf{G}_F = \mathbf{F}, \tag{139}$$

$$\mathbf{G}_F^T\mathbf{G}_F = \mathbf{I}_k. \tag{140}$$

The matrix $\mathbf{G}_F$ is a sparse matrix with zeros everywhere, except for those places where we are "selecting" the right column vectors,

$$G_{F,bc} = \delta_{bb(c)}, \tag{141}$$

where as above $b \in [1, m]$ is the index that selects columns of $\mathbf{T}$, $c \in [1, k]$ is the index that selects columns of $\mathbf{F}$, and $\delta_{bb(c)}$ is a Kronecker delta that selects when a column of $\mathbf{T}$ is also a column of $\mathbf{F}$. Let us now insert the projector into the detection statistic,

$$\begin{aligned} D(y, \theta) &= y_3^T\mathbf{L}_B^{-1}\mathbf{G}_T^T\mathbf{L}_\mathrm{W}^{-1}\mathbf{T}\mathbf{P}_F\Delta\mathbf{B}\mathbf{P}_F^T\mathbf{T}^T\mathbf{L}_\mathrm{W}^{-T}\mathbf{G}_T\mathbf{L}_B^{-T}y_3/\mathcal{n} \\ &= y_3^T\mathbf{L}_B^{-1}\mathbf{G}_T^T\mathbf{L}_\mathrm{W}^{-1}\mathbf{T}\mathbf{G}_F\Delta\mathbf{\Phi}\mathbf{G}_F^T\mathbf{T}^T\mathbf{L}_\mathrm{W}^{-T}\mathbf{G}_T\mathbf{L}_B^{-T}y_3/\mathcal{n}, \end{aligned} \tag{142}$$

where we have defined $\Delta\Phi = \mathbf{G}_F^T\Delta B\mathbf{G}_F$. The projector $\mathbf{P}_F$ does not commute with $\mathbf{A} = \mathbf{L}_B^{-1}\mathbf{G}_T^T\mathbf{L}_\mathrm{W}^{-1}\mathbf{T}$, so unlike before we cannot just "move" the projector toward $y_3$ and project the data onto some subspace to compress the system even further. But, we know that the rank of the system cannot be higher than the rank of $\mathbf{P}_F$. The thin SVD will therefore be able to give us the projection basis we need: $\mathbf{A}\mathbf{G}_F = \mathbf{G}_A\mathbf{\Sigma}\mathbf{V}^T$. The interpretation is that $\mathbf{V}$ represents the basis of our system in the column space of $\mathbf{F}$, whereas $\mathbf{G}_A$ corresponds to those degrees of freedom in the basis of $y_3$. The projection matrix $\mathbf{P}_A = \mathbf{G}_A\mathbf{G}_A^T$ can therefore be inserted in the detection statistic,

$$D(y, \theta) = y_3^T\mathbf{P}_A\mathbf{L}_B^{-1}\mathbf{G}_T^T\mathbf{L}_\mathrm{W}^{-1}\mathbf{F}\Delta\mathbf{\Phi}\mathbf{F}^T\mathbf{L}_\mathrm{W}^{-T}\mathbf{G}_T\mathbf{L}_B^{-T}\mathbf{P}_A y_3/\mathcal{n}. \tag{143}$$

With this, we find the last compression step, transformation 4,

$$y_4 = \mathbf{G}_A^T y_3 = \mathbf{G}_A^T\mathbf{L}_B^{-1}\mathbf{G}_T^T\mathbf{L}_\mathrm{W}^{-1}y, \tag{144}$$

$$\mathbf{N}_4 = \langle y_4 y_4^T\rangle_{H_0} = \mathbf{I}_k, \tag{145}$$

$$\begin{aligned} D(y, \theta) &= y_4^T\mathbf{G}_A^T\mathbf{L}_B^{-1}\mathbf{G}_T^T\mathbf{L}_\mathrm{W}^{-1}\mathbf{F}\Delta\mathbf{\Phi}\mathbf{F}^T\mathbf{L}_\mathrm{W}^{-T}\mathbf{G}_T\mathbf{L}_B^{-T}\mathbf{G}_A y_4/\mathcal{n} \\ &= \frac{y_4^T\mathbf{R}\Delta\mathbf{\Phi}\mathbf{R}^T y_4}{\sqrt{2\mathrm{Tr}[\mathbf{R}\Delta\mathbf{\Phi}\mathbf{R}^T\mathbf{R}\Delta\mathbf{\Phi}\mathbf{R}^T]}}, \\ \mathbf{Q}_4 &= \frac{\mathbf{R}\Delta\mathbf{\Phi}\mathbf{R}^T}{\sqrt{2\mathrm{Tr}[\mathbf{R}\Delta\mathbf{\Phi}\mathbf{R}^T\mathbf{R}\Delta\mathbf{\Phi}\mathbf{R}^T]}}, \end{aligned} \tag{146}$$

where on the last line we have defined $\mathbf{R} = \mathbf{G}_A^T\mathbf{L}_B^{-1}\mathbf{G}_T^T\mathbf{L}_\mathrm{W}^{-1}\mathbf{F}$. The transformation matrices $\mathbf{R}$ and data vectors $y_4$ can be calculated per pulsar, which is not computationally expensive. That also means that the normalization trace in the denominator can be evaluated without expensive large matrix products.

We now have our reduced-size quadratic filter $\tilde{\mathbf{Q}} = \mathbf{Q}_4$ and our reduced-size data $\tilde{y} = y_4$. Under $H_0$ the data $\tilde{y}$ is a random variable with uncorrelated elements distributed as a standard Gaussian. That means we can find the distribution of $D(\tilde{y}, \theta)$ by solving for the eigenvalues of $\tilde{\mathbf{Q}}$. The detection statistic itself is just $\tilde{y}^T\tilde{\mathbf{Q}}\,\tilde{y}$. Noteworthy properties of $\tilde{y}$ and $\tilde{\mathbf{Q}}$ are,

$$\tilde{y} \sim \mathcal{N}(0, \mathbf{I}_k), \tag{147}$$

$$\mathrm{Tr}[\tilde{\mathbf{Q}}] = 0, \tag{148}$$

$$\mathrm{Tr}[\tilde{\mathbf{Q}}^2] = \frac{1}{2}. \tag{149}$$

The zero-trace property of $\tilde{\mathbf{Q}}$ indicates that our detection statistic is zero in expectation under $H_0$. The last identity indicates that our detection statistic is normalized to 1, meaning it has unit variance. Both traces over $\tilde{\mathbf{Q}}$ also hold for its eigenvalues. Again here, the factor of two (or one-half) comes from the fact that we are dealing with real-valued random variables here, rather than complex-valued variables as in the rest of this manuscript.

The above prescription of the reduced-size quadratic filter can also be used in conjunction with the weighted uniform Dirichlet distribution, which then gives a semi-analytic route to calculating the scrambling $p$-values. However, as noted in earlier sections, we do not recommend that approach and instead the generalized $\chi^2$ distribution should be used.

### D. Rigorous PTA $p$-values

Using the compressed data and quadratic filter of the detection statistic we can formulate the statistically rigorous approach to calculate Bayesian and Frequentist $p$-value. For a Frequentist $p$-value we need a means to sample representative realizations of data from $H_0$. This is model dependent, and can be done with simulations or other means. Then,

(i) Estimate the model parameters $\hat{\theta}$ from the real data
(ii) Use those parameters and real data to evaluate $D(z^{\mathrm{obs}}, \hat{\theta})$
(iii) Simulate realizations of data $z \sim p(z|H_0)$
(iv) Estimate model parameters $\tilde{\theta}(z)$ for each realization of $z$
(v) Construct reduced data $\tilde{y}$ and filter $\tilde{\mathbf{Q}}$ for each realization
(vi) Use the generalized $\chi^2$ distribution to find a $p$-value of the real data for these estimated model parameters

The Frequentist $p$-value is the average of the $p$-values that were found over all realizations of $z$.

A Bayesian $p$-value is similarly calculated, but now the model parameters and *data replications* are drawn from the posterior predictive $p(z, \theta|z^{\mathrm{obs}}, H_0)$, as described by Vallisneri *et al.* [11] and Agazie *et al.* [46]. The recipe is then,

(i) Iterate over posterior model parameters $\theta$ from an MCMC analysis
(ii) Construct the reduced data $\chi$ and quadratic filter $\tilde{\mathbf{Q}}$
(iii) Calculate the detection statistic of the real data and
(iv) Use the generalized $\chi^2$ distribution to find a $p$-value for these model parameters

The Bayesian $p$-value is then the average of all $p$-values found for all MCMC samples. We have applied the above recipe to the NANOGrav 15-year dataset, and we found the same $p$-value as Agazie *et al.* [46].

## X. CONCLUSIONS

We investigated the optimal detection statistic for the detection of a stochastic background of GWs in PTA experiments. Various methods to calculate $p$-values are studied and many novel analytical results are presented. In the context of quadratic detection statistics, we have derived the general form of scrambling operations under which the null hypothesis $H_0$ remains invariant, and we have proved that scrambling methods indeed cancel correlations in the data. The distribution of the detection statistic under scrambling operations turns out to be a weighted uniform Dirichlet distribution, which is a model-dependent expression. We conclude that scrambling is not equivalent to drawing data from $H_0$, which in hindsight can also be seen from the fact that the estimated model parameters do not change when scrambling the data (which *would* happen under $H_0$).

We investigate rigorous Bayesian and Frequentist $p$-value methods for quadratic detection statistics. We arrive at the posterior predictive $p$-value that was introduced by [11] and first used by [46] as the correct Bayesian approach to calculating $p$-values. To make such analyses practical, we derive efficient expressions to carry out the generalized $\chi^2$ calculations introduced by [10]. Prior to our presentation such calculations required expensive numerical simulations. We find full consistency with published posterior predictive $p$-values in the literature when reanalyzing the data with our expressions. The final recommendation of this paper is to replace the use of scrambling operations with the calculation of a posterior predictive $p$-value.

## ACKNOWLEDGMENTS

We thank Joe Romano for in-depth discussions and many comments on an early draft of this manuscript. The many discussions with Bruce Allen have been very useful in the shaping of the direction of this paper. We thank Jeff Hazboun, Michele Vallisneri, Kyle Gersbach, and Pat Meyers for their assistance in porting the code for the generalized chi-square distribution and cross-checking our results with established methods. Valentina di Marco and Stephen Taylor are thanked for feedback on a draft of this manuscript. Lastly, we thank Wang-Wei Yu for feedback regarding the clarity of the discussion of the Weingarten function.

## DATA AVAILABILITY

The data that support the findings of this article are openly available [57].

## APPENDIX A: COMPLEX MULTIVARIATE NORMAL RANDOM VARIABLES

We use multivariate normal random variables. Usually results regarding multivariate normal distributions are only given for real variables, and subtleties can be overlooked. We therefore give some basic results here explicitly. We define complex multivariate random variables as,

$$z \sim \mathcal{N}^{\mathbb{C}}(0, \mathbf{C}), \tag{A1}$$

$$z = \frac{x + Jy}{\sqrt{2}}, \tag{A2}$$

$$x \sim \mathcal{N}(0, \mathbf{C}), \tag{A3}$$

$$y \sim \mathcal{N}(0, \mathbf{C}), \tag{A4}$$

where $J$ is the imaginary number with $J^2 = -1$, and $\mathcal{N}^{\mathbb{C}}$ indicates a complex multivariate random variable. The expectation value of a complex random variable like above is zero,

$$\langle z_a \rangle = 0. \tag{A5}$$

This makes sense: we had defined our multivariate normal distributions as zero mean. The quadratic terms are,

$$\langle z_a^* z_b \rangle = \langle (x - Jy)_a (x + Jy)_b \rangle = C_{ab}. \tag{A6}$$

The quartic terms can be found using Isserlis's theorem [40],

$$\langle z_a^* z_b z_c^* z_d \rangle = C_{ab} C_{cd} + C_{ad} C_{bc}. \tag{A7}$$

This result is useful when we need to calculate the expected value of a detection statistic, which requires sums like,

$$\sum_{abcd} Q_{ab} Q_{cd} \langle z_a^* z_b z_c^* z_d \rangle = \mathrm{Tr}(\mathbf{QC})^2 + \mathrm{Tr}(\mathbf{QCQC}). \tag{A8}$$

Note that for real-valued random variables Isserlis' theorem will contain one extra term, leading to a factor of two in the normalization.

## APPENDIX B: HELLINGS & DOWNS SKY SCRAMBLE DISTRIBUTION

Sky scrambling is best interpreted as a transformation of the quadratic filter $\mathbf{Q}$. If the scrambles are drawn uniformly from the sky, that means that for every pulsar pair the angle between them $\gamma_{ab}$ is distributed as,

$$\cos(\gamma_{ab}) \sim \mathrm{Uniform}(-1, 1). \tag{B1}$$

We now ask what the density of H&D correlations becomes if $\gamma_{ab}$ is distributed like Eq. (B1). Remember that the H&D correlations are defined as:

$$\xi_{ab} = \frac{(1 - \cos\gamma_{ab})}{2}, \tag{B2}$$

$$\mu(\gamma_{ab}) = \frac{3\xi_{ab} \log \xi_{ab}}{2} - \frac{1}{4}\xi_{ab} + \frac{1}{2}(1 + \delta_{ab}). \tag{B3}$$

This means that $\xi_{ab} \sim \mathrm{Uniform}(0, 1)$. The usual approach to finding a density would be to invert Eq. (B2), and use the derivative of the inverse. That is difficult to do directly, because Eq. (B2) is a transcendental equation, so it cannot be inverted using elementary operations. Fortunately, we can avoid the inverse. We drop the pulsar term,

$$\frac{\mathrm{d}\mu}{\mathrm{d}\xi} = \frac{3 \log \xi}{2} + \frac{5}{4} \tag{B4}$$

$$\frac{\mathrm{d}\xi}{\mathrm{d}\mu} = \left(\frac{\mathrm{d}\mu}{\mathrm{d}\xi}\right)^{-1} = \left(\frac{3 \log \xi}{2} + \frac{5}{4}\right)^{-1}. \tag{B5}$$

The last line is written as function of $\xi$, not $\mu$, so we still cannot do the inverse. In fact, $\xi(\mu)$ is double valued, and we need to take some care when continuing. It is necessary to break up the curve $\xi(\mu)$ in two parts: $\xi < \xi_{\mathrm{min}}$ and $\xi > \xi_{\mathrm{min}}$, with $\xi_{\mathrm{min}}$ the value of $\xi$ at the minimum of the H&D curve $\mu(\xi)$. In the left panel of Fig. 3 we show the curve $\xi(\mu)$, and in the middle panel we show the curve of $\mathrm{d}\xi/\mathrm{d}\mu(\mu)$. The derivative $\mathrm{d}\xi/\mathrm{d}\mu(\mu)$ is related to the density $p(\mu)$ we want to characterize: we need to sum the derivative corrected for the direction of $\xi$ along the path. The points $S$ and $F$ represent the start and the finish of our path, and the top $T$ and the bottom $B$ actually represent the same point (the minimum of the H&D curve), which here both lie at $\mathrm{d}\xi/\mathrm{d}\mu(\mu) = \pm\infty$. The density $p(\mu)$ is the sum of the two components (corrected for sign) plotted in the middle panel. We have shown that density in the right panel. Note the discontinuity at $\mu = 1/4$, which is due to the curve stopping at $(\xi, \mu) = (1, 1/4)$. This discontinuity was the reason we created Fig. 3, because we noticed the discontinuity when inspecting the distribution of the correlation matrix under sky scrambling operations. This density shows, for isotropically distributed pulsars, how likely it is to have a single pulsar pair with a specific H&D correlation. We can also calculate the average correlation using the above equations,

$$\langle \mu \rangle = \int \mathrm{d}\mu \, p(\mu) \mu = \int_S^F \mathrm{d}\mu \frac{\mathrm{d}\xi}{\mathrm{d}\mu} \mu = 0. \tag{B6}$$

Equation (B6) can be evaluated analytically, and we find that it is zero. This is consistent with what we found in Sec. IV E 1. We can similarly find the spread of the H&D correlations under $\mathrm{d}\xi$,

$$\langle \mu^2 \rangle = \int_0^1 \mathrm{d}\xi \, \mu^2 = \frac{1}{48}. \tag{B7}$$

The above two equations are also possible to calculate using coordinates in $\cos\gamma$.

## APPENDIX C: RELEVANT STATISTICAL DISTRIBUTIONS

For certain results in the upcoming appendixes we need some basic identities of common distributions: the gamma distribution and the Dirichlet distribution. We present and discuss those here.

### 1. The gamma distribution

The gamma distribution is defined as,

$$p(x|\theta, \alpha) = \frac{x^{\alpha-1} e^{-x/\theta}}{\theta^{\alpha} \Gamma(\alpha)} \tag{C1}$$

where $\alpha$ is called the shape parameter, and $\theta$ is called the scale parameter. We can equivalently write,

$$x \sim \mathrm{Gamma}(\alpha, \theta). \tag{C2}$$

We always write $\Gamma(x)$ for the gamma function (the common generalization of the factorial). The exponential distribution is a special case of the gamma distribution with $\alpha = 1$, and $1/\theta$ would then be the rate parameter. The chi-squared distribution with $k$ degrees of freedom is also a special case of the gamma distribution, with $\alpha = k/2$ and $\theta = 2$.

The sum of multiple gamma-distributed random variables is itself also a gamma-distributed random variable if all gamma distributions have the same scale parameter $\theta$,

$$x_i \sim \mathrm{Gamma}(\alpha_i, \theta), \tag{C3}$$

$$x = \sum_i x_i, \tag{C4}$$

$$\alpha_0 = \sum_i \alpha_i, \tag{C5}$$

$$x \sim \mathrm{Gamma}(\alpha_0, \theta). \tag{C6}$$

### 2. Properties of the Dirichlet distribution

The Dirichlet distribution is a common and well-studied distribution in probability and statistics. Here we list several properties. The Dirichlet distribution of order $M \geq 2$ with *concentration* parameters $\alpha_1, \alpha_2, \ldots, \alpha_M > 0$ has a probability density function given by,

$$p(x_1, x_2, \ldots, x_M | \alpha_1, \alpha_2, \ldots, \alpha_M) = \frac{1}{B(\alpha)} \prod_{i=1}^{M} x_i^{\alpha_i - 1}, \tag{C7}$$

where $x_i \in [0, 1]$ for all $i = 1, \ldots, M$, with the constraint that $\sum_{i=1}^{M} x_i = 1$. This means that $x$ lies on the standard $M-1$-dimensional simplex. The normalization constant $B(\alpha)$ is the beta function, defined in terms of the gamma function $\Gamma(x)$,

$$B(\alpha) = \frac{\prod_{i=1}^{M} \Gamma(\alpha_i)}{\Gamma(\alpha_0)}, \tag{C8}$$

where we use the convention in this paper that $\alpha_0 = \sum_{i=1}^{M} \alpha_i$. We write:

$$x \sim \mathrm{Dirichlet}(\alpha_1, \alpha_2, \ldots, \alpha_M). \tag{C9}$$

The Dirichlet distribution has statistical properties,

$$\mathbb{E}[x_i] = \frac{\alpha_i}{\alpha_0}, \tag{C10}$$

$$\mathrm{Var}[x_i] = \frac{\alpha_i(\alpha_0 - \alpha_i)}{\alpha_0^2(\alpha_0 + 1)}, \tag{C11}$$

$$\mathrm{Cov}(x_i, x_j) = \frac{-\alpha_i \alpha_j}{\alpha_0^2(\alpha_0 + 1)}. \tag{C12}$$

For the uniform Dirichlet distribution $\mathrm{Dirichlet}(1, 1, \ldots, 1)$ this becomes,

$$\mathbb{E}[x_i] = \frac{1}{M}, \tag{C13}$$

$$\mathrm{Var}[x_i] = \frac{(M-1)}{M^2(M+1)}, \tag{C14}$$

$$\mathrm{Cov}(x_i, x_j) = \frac{-1}{M^2(M+1)}. \tag{C15}$$

### 3. Dirichlet as a combination of gamma distributions

It is possible to express the Dirichlet distribution as a combination of gamma-distributed random variables. Let $g_i$ be $M$ gamma-distributed random variables with scale parameter $\theta$ and shape parameters $\alpha_i$ with $i = 1, 2, \ldots, M$. We can then write the joint probability density function as,

$$p(g_1, g_2, \ldots, g_M | \theta, \alpha) = \prod_{i=1}^{M} \frac{g_i^{\alpha_i - 1} e^{-g_i/\theta}}{\theta_i^{\alpha} \Gamma(\alpha_i)}. \tag{C16}$$

We will now introduce a change of coordinates,

$$g \coloneqq \sum_{i=1}^{M} g_i, \tag{C17}$$

$$w_i \coloneqq \frac{g_i}{g}. \tag{C18}$$

This means that $\sum_i w_i = 1$. Our goal is now to show that $w$ is distributed according to a Dirichlet distribution with concentration parameters $\alpha_i$. We therefore start with a coordinate transformation,

$$(g_1, g_2, \ldots, g_M) \rightarrow (w_1, w_2, \ldots, w_{M-1}, g). \tag{C19}$$

The partial derivatives of this transformation are,

$$\frac{\partial g_i}{\partial w_j} = \delta_{ij} g, \tag{C20}$$

$$\frac{\partial g_i}{\partial g} = w_i \quad \text{for } i < M, \tag{C21}$$

$$\frac{\partial g_M}{\partial w_i} = -g, \tag{C22}$$

$$\frac{\partial g_M}{\partial g} = 1, \tag{C23}$$

where we have used that $w_M = 1 - \sum_{i=1}^{M-1} w_i$. The Jacobian of the transformation is therefore,

$$J = \begin{vmatrix} g & 0 & 0 & \ldots & -g \\ 0 & g & 0 & & -g \\ 0 & 0 & g & & -g \\ \vdots & & & \ddots & \vdots \\ w_1 & w_2 & w_3 & \ldots & 1 \end{vmatrix} = g^{M-1} w_M. \tag{C24}$$

This is most straightforwardly found using the identity,

$$\begin{vmatrix} \mathbf{A} & b \\ c^T & d \end{vmatrix} = \det \mathbf{A} \cdot (d - c^T \mathbf{A}^{-1} b). \tag{C25}$$

We can now rewrite the probability density of our joint-gamma distribution in terms of $(w_1, \ldots, w_{M-1}, g)$,

$$\begin{aligned} p(w_1, \ldots, w_{M-1}, g) \\ &= \frac{g^{\sum_{i=1}^M (\alpha_i - 1)} g^{M-1} e^{-g/\theta}}{\prod_{i=1}^M \Gamma(\alpha_i)} \frac{w_M^{\alpha_M} \prod_{i=1}^{M-1} w_i^{\alpha_i - 1}}{\theta^{\sum_{i=1}^M \alpha_i}} \\ &= \frac{g^{\alpha_0 - 1} e^{-g/\theta}}{\prod_{i=1}^M \Gamma(\alpha_i)} \frac{w_M^{\alpha_M} \prod_{i=1}^{M-1} w_i^{\alpha_i - 1}}{\theta^{\alpha_0}}, \end{aligned} \tag{C26}$$

where we defined $\alpha_0 = \sum_{i=1}^M \alpha_i$ to clear up notation. At this point we observe that the parameter $g$ that represents the sum $g = \sum_i g_i$ is a normalization parameter that does not occur in the Dirichlet distribution. Our goal is to show that the $w_i$ follows a Dirichlet distribution. We therefore have to marginalize out $g$ from the joint distribution. The sum parameter $g$ is the sum of $M$ gamma-distributed random variables, which is itself then also a gamma-distributed random variable. Indeed, we recognize the gamma-distribution in Eq. (C26). We can therefore make the following substitution,

$$\int_0^\infty \mathrm{d}g \, g^{\alpha_0 - 1} e^{-g/\theta} = \theta^{\alpha_0} \Gamma(\alpha_0). \tag{C27}$$

This gives us the following marginal distribution:

$$p(w_1, \ldots, w_{M-1}) = \frac{\Gamma(\alpha_0)}{\prod_{i=1}^M \Gamma(\alpha_i)} \left( \prod_{i=1}^{M-1} w_i^{\alpha_i - 1} \right) w_M^{\alpha_M}. \tag{C28}$$

This is almost identical to the Dirichlet distribution,

$$p(w_1, \ldots, w_M) = \frac{\Gamma(\alpha_0)}{\prod_{i=1}^M \Gamma(\alpha_i)} \prod_{i=1}^M w_i^{\alpha_i - 1}. \tag{C29}$$

The difference between Eq. (C28) and Eq. (C29) is that the former is the distribution represented in terms of $(w_1, \ldots, w_{M-1})$ rather than $(w_1, \ldots, w_M)$. However, $w_M = 1 - \sum_{i=1}^{M-1} w_i$, so with the marginal distribution of Eq. (C28) the distribution is fully specified. The difference is just a matter of normalization with respect to a different domain. With this, we have proved that the Dirichlet distribution can be written as a product of $M$ independent gamma-distributed random variables.

### a. Gamma distribution from Dirichlet

We may also reverse the derivation above. Let $w$ be a Dirichlet-distributed $M$-dimensional random variable with concentration parameters $\alpha_i$, and let $u$ be a gamma-distributed random variable with scale parameter $\theta$ and shape parameter $\alpha_0 = \sum_i \alpha_i$,

$$w \sim \text{Dirichlet}(\alpha_1, \alpha_2, \ldots, \alpha_M), \tag{C30}$$

$$u \sim \text{Gamma}(\alpha_0, \theta). \tag{C31}$$

$$\tag{C32}$$

Then we have,

$$u_i = u w_i, \tag{C33}$$

$$u_i \sim \text{Gamma}(\alpha_i, \theta). \tag{C34}$$

The resulting $u_i$ are therefore also gamma distributed, with the scale parameter $\theta$ from the gamma distribution for $u$, and the shape parameters $\alpha_i$ equal to the concentration parameters $\alpha_i$ of the Dirichlet distribution. This relationship shows a deep connection between the gamma distribution and the Dirichlet distribution.

## APPENDIX D: RANDOM UNITARY MATRIX IDENTITIES

The complete set of linear scrambles we propose are based on the group of unitary matrices. A natural way to define a uniform probability distribution over the group of unitary matrices is provided by the Haar measure [58]. The unitary scrambles are thus random draws of Haar-distributed unitary matrices.

To calculate expectation values of the detection statistic and related quantities, we use results from Collins and Śniady [59], which is a standard reference in the field of random matrix theory. This reference includes discussions of integration with respect to the Haar measure under the unitary group. We often find such mathematical works hard to parse, so we review the results we need in this appendix. Once we understood the notation below, we found the original reference easier to understand.

We begin by defining the unitary group $U(M)$ as the group of $(M\times M)$ unitary matrices $\mathbf{S}$, where a matrix $\mathbf{S}$ is unitary if $\mathbf{S}^\dagger\mathbf{S}=\mathbf{S}\mathbf{S}^\dagger=\mathbf{I}$, with $\mathbf{I}$ the identity matrix. The Haar measure $\mathrm{d}\nu(S)$ is the unique, translation-invariant measure on the unitary group, which defines a uniform probability density over $U(M)$. More formally, the Haar measure over $U(M)$ is defined by,

$$\int_{U(M)} f(\mathbf{U})\mathrm{d}\mu(U) = \int_{U(M)} f(\mathbf{V}\mathbf{U})\mathrm{d}\mu(U) = \int_{U(M)} f(\mathbf{U}\mathbf{V})\mathrm{d}\mu(U), \tag{D1}$$

for any unitary matrix $\mathbf{V}\in U(M)$ and any measurable function $f(\mathbf{U})$. This invariance property ensures that transformations by Haar-distributed unitary matrices do not change the measure. If $\mathbf{S}$ is a Haar-distributed random unitary matrix, we write, $\mathbf{S}\sim\mathcal{H}U(M)$.

### 1. Expectations under the unitary group

The definition of Eq. (D1) ensures that the unitary matrices are uniformly distributed on $M^2$-dimensional subspace of all $(2M)^2$-dimensional $(M\times M)$ invertible complex matrices (this general Lie-group of matrices can be referred to as the Ginibre ensemble: denoted as $\mathrm{GL}(M,\mathbb{C})$). This means that, if $\hat{e}_1=(1,0,\ldots,0)^T$ is the first unit basis vector in $\mathbb{C}^M$, then $\mathbf{U}\hat{e}_1$ is uniformly distributed on the complex $M$ sphere if $\mathbf{U}$ is Haar distributed. This uniformity property is powerful when one is interested in expectation values. In this appendix we derive various useful quantitites that allow us to calculate expectations under Haar-distributed random unitary matrices. We start with,

$$\mathbb{E}_{U(M)}[U_{ia}]=0, \tag{D2}$$

this follows trivially from the above that $\mathbf{U}\hat{e}_1$ is uniformly distributed on the complex $M$ sphere. In general, combinations of $U$ with an odd number of $U_{ij}$ are zero in expectation. We now consider the first nontrivial case,

$$\mathbb{E}_{U(M)}[U_{ia}U^*_{jb}]=\alpha\delta_{ab}\delta_{ij}. \tag{D3}$$

Because $\mathbf{U}$ is unitary and Haar distributed, the expectation has to be proportional to $\delta_{ij}\delta_{ab}$: this follows from invariance under index relabelings and the uniformity/isotropy of the group. What is left is to determine the proportionality constant $\alpha$. We obtain that by observing that for any unitary matrix, we have,

$$\sum_{a=1}^{M} U_{ai}U^*_{aj}=\delta_{ij}. \tag{D4}$$

If we substitute that into Eq. (D3), we obtain,

$$\sum_{a=1}^{M}\mathbb{E}_{U(M)}[U_{ai}U^*_{aj}]=\delta_{ij}, \tag{D5}$$

$$\sum_{a=1}^{M}\alpha\delta_{aa}\delta_{ij}=\delta_{ij}, \tag{D6}$$

$$\alpha=\frac{1}{M}, \tag{D7}$$

which means that,

$$\mathbb{E}_{U(M)}[U_{ia}U^*_{jb}]=\frac{1}{M}\delta_{ab}\delta_{ij}. \tag{D8}$$

Let $\mathbf{G}$ be an $(M\times M)$ complex matrix; then we may be interested in the expectation of all Haar-distributed similarity transformations of $\mathbf{G}$: $\mathbf{S}\mathbf{G}\mathbf{S}^\dagger$. We first note that $\mathbf{G}$ can be diagonalized by some similarity transformation, and following the definition of the Haar measure in Eq. (D1) this means that we may therefore replace $\mathbf{G}$ with $\mathbf{\Lambda}$, where $\Lambda_{ij}=g_i\delta_{ij}$ with $g_i$ the eigenvalues of $\mathbf{G}$. Then,

$$\mathbb{E}_{U(M)}\left[\sum_{ij}U_{ai}G_{ij}U^*_{bj}\right]=\mathbb{E}_{U(M)}\left[\sum_{ij}U_{ai}g_i\delta_{ij}U^*_{bj}\right], \tag{D9}$$

$$=\left(\sum_{i}\frac{1}{M}g_i\delta_{ij}\right)\delta_{ab}. \tag{D10}$$

The term in the parentheses on the last line is just the average of all eigenvalues of $\mathbf{G}$.

### 2. Weingarten function

Similar to what we just did for expectation values of second-order combinations of $\mathbf{U}$ we can also calculate higher-order combinations. For the general unitary group, this involves combinatorics that can get complicated. The Weingarten function $\mathrm{Wg}(\sigma, M)$ is defined to make that easier, so we briefly discuss essential results here. We are only going to consider fourth order combinations of matrices of the type,

$$\mathbb{E}_{U(M)}[S_{ia}S^*_{jb}S_{kc}S^*_{ld}] = \sum_p \alpha_p \delta_{i'j'}\delta_{k'l'}\delta_{a'b'}\delta_{c'd'} \tag{D11}$$

where the summation is over all permutations of the indices $(i, j, k, l, a, b, c, d)$, and the prime on the indices means these indices have been permutated by $p$,

$$p:(i, j, k, l, a, b, c, d) \rightarrow (i', j', k', l', a', b', c', d'). \tag{D12}$$

For combinations of four unitary matrices as in Eq. (D11) we have eight indices, meaning that the general expression would already have 8! different permutations. Fortunately, the expression greatly simplifies for the product of unitary matrices, and the Weingarten function helps to keep track of everything.

The results in Collins and Śniady [59] use the general Weingarten function, which can be thought of as an inverse of a matrix built from permutations of the symmetric group $S_k$. For our purposes, we only need elements from $S_2 = \{e, t\}$, which are the identity permutation $e$ of two elements (leave both as-is) and the transposition permutation $t$ (swap the indices). We found the notation in the literature slightly confusing, so we use the following notation. Our "permutations" are actually functions that reorder (permutate) the input and then pick/return the first one. These functions we denote with $\sigma(i, j)$ and $\tau(i, j)$, where $\sigma, \tau \in \{e, t\}$. If $\sigma = e$, then $\sigma(i, j) = i$. If $\sigma = t$, then $\sigma(i, j) = j$. This notation is close to what is used in the literature, but slightly more explicit.

The simplified Weingarten function on $S_2$ is given by,

$$\mathrm{Wg}(e, M) = \frac{1}{M^2 - 1}, \quad \mathrm{Wg}(t, M) = \frac{-1}{M(M^2 - 1)}. \tag{D13}$$

Using these expressions and notation it becomes easier to read the results from Collins and Śniady [59]. Now we can give the expectation value for the product of two and four matrix elements of a unitary matrix $\mathbf{S}$ under the Haar measure. For two elements

$$\mathbb{E}_{U(M)}[S_{ia}S^*_{jb}] = \frac{1}{M}\delta_{ij}\delta_{ab}, \tag{D14}$$

which is the result we also found in Eq. (D8). For four matrix elements we have [ [59] Eq. (11)],

$$\mathbb{E}U_{(M)}[S_{ia}S^*_{jb}S_{kc}S^*_{ld}] = \sum_{\sigma,\tau \in S_2} \delta_{i\sigma(j,l)}\delta_{k\sigma(l,j)}\delta_{a\tau(b,d)}\delta_{c\tau(d,b)}\mathrm{Wg}(\sigma\tau^{-1}, M). \tag{D15}$$

We see that we only have to consider four terms in our sum. Both $\sigma$ and $\tau$ can only be one of two permutations. We explicitly work out all the terms here, including the values of the Weingarten function,

$$\sigma = e, \tau = e{:}\ \frac{1}{M^2 - 1}\delta_{ij}\delta_{kl}\delta_{ab}\delta cd, \tag{D16}$$

$$\sigma = t, \tau = t{:}\ \frac{1}{M^2 - 1}\delta_{il}\delta_{kj}\delta_{ad}\delta cb, \tag{D17}$$

$$\sigma = t, \tau = e{:}\ \frac{-1}{M(M^2 - 1)}\delta_{il}\delta_{kj}\delta_{ab}\delta cd, \tag{D18}$$

$$\sigma = e, \tau = t{:}\ \frac{-1}{M(M^2 - 1)}\delta_{ij}\delta_{kl}\delta_{ad}\delta cb. \tag{D19}$$

This gives,

$$\begin{aligned}\mathbb{E}_{U(M)}&[S_{ia}S^*_{jb}S_{kc}S^*_{ld}] \\ &= \frac{1}{M^2 - 1}(\delta_{ij}\delta_{kl}\delta_{ab}\delta_{cd} + \delta_{il}\delta_{jk}\delta_{ad}\delta_{cb}) \\ &\quad - \frac{1}{M(M^2 - 1)}(\delta_{ij}\delta_{kl}\delta_{ad}\delta_{cb} + \delta_{il}\delta_{jk}\delta_{ab}\delta_{cd}).\end{aligned} \tag{D20}$$

Similar but more cumbersome results are available as well for higher-order combinations. These equations show that the expectation value is proportional to the product of Kronecker deltas, scaled by functions of $M$. They reflect the inherent symmetry of the unitary group $U(M)$ and the uniformity of the Haar measure. In Sec. IV we used an intuitive argument similar to Eq. (D14).

### 3. Scrambles under the unitary group

When discussing the effects of transformations of the unitary group on the detection statistic, we are calculating products of vectors with the weights $Q_{ij}$, where $\mathbf{Q}$ is a real symmetric traceless $(M \times M)$ matrix. For the average detection statistic, we then find,

$$\mathbb{E}_{U(M)}[S(\mathbf{U}z)] = \int \mathrm{d}\nu(S) z^\dagger \mathbf{S}^\dagger \mathbf{Q}\mathbf{S} z \tag{D21}$$

$$= \int \mathrm{d}\nu(S) \sum_{ijab} z_i z_j^\dagger S_{ai}^\dagger S_{bj} \tag{D22}$$

$$= \sum_{ijab} \frac{1}{M} \delta_{ij}\delta_{ab} z_i z_j^\dagger Q_{ab} = 0, \tag{D23}$$

where on the last line we used that $\mathbf{Q}$ is a traceless matrix, and we used Eq. (D14). The variance of the detection statistic can be found going through the same motions for higher order combinations of products of matrices of the unitary group,

$$\mathbb{E}_{U(M)}[D(\mathbf{U}z)^2] = \int \mathrm{d}\nu(S) (z^\dagger \mathbf{S}^\dagger \mathbf{Q}\mathbf{S} z)^2$$
$$= \int \mathrm{d}\nu(S) \sum_{\substack{abcd\\ijkl}} z_i z_j^\dagger z_k z_l^\dagger S_{ai}^\dagger S_{bj} S_{ck}^\dagger S_{dl}. \tag{D24}$$

We now substitute $z = |z|\mathbf{V}\hat{e}_m$, where $\mathbf{V} \in U(M)$ is some unitary matrix, $|z|$ is the complex amplitude of $z$, and $\hat{e}_m$ is the $m$th unit basis vector with elements $(\hat{e}_m)_i = \delta_{im}$. We are allowed to choose an arbitrary index $m$, with which we find,

$$\mathbb{E}_{U(M)}[D(\mathbf{U}z)^2] = |z|^4 \mathrm{Tr}(\mathbb{E}_{U(M)}[\mathbf{U}\hat{e}_m\hat{e}_m^\dagger \mathbf{U}^\dagger \mathbf{Q}\mathbf{U}\hat{e}_m\hat{e}_m^\dagger \mathbf{U}^\dagger \mathbf{Q}]). \tag{D25}$$

Symmetry allows us to also just sum over *all* values of $m$, and divide the total answer by $M$. This gives,

$$\begin{aligned}
\mathbb{E}_{U(M)}[D(\mathbf{U}z)^2] &= |z|^4 \mathrm{Tr}(\mathbb{E}_{U(M)}[\mathbf{U}\hat{e}_m\hat{e}_m^\dagger \mathbf{U}^\dagger \mathbf{Q}\mathbf{U}\hat{e}_m\hat{e}_m^\dagger \mathbf{U}^\dagger \mathbf{Q}]) \\
&= |z|^4 \frac{1}{M} \sum_{m=1}^{M} \mathrm{Tr}(\mathbb{E}_{U(M)}[\mathbf{U}\hat{e}_m\hat{e}_m^\dagger \mathbf{U}^\dagger \mathbf{Q}\mathbf{U}\hat{e}_m\hat{e}_m^\dagger \mathbf{U}^\dagger \mathbf{Q}]) \\
&= |z|^4 \frac{1}{M} \sum_{\substack{abcd\\ijkl}} \left( \frac{1}{M^2-1} (\delta_{ij}\delta_{kl}\delta_{ab}\delta_{cd} + \delta_{il}\delta_{jk}\delta_{ad}\delta_{cb}) - \frac{1}{M(M^2-1)} (\delta_{ij}\delta_{kl}\delta_{ad}\delta_{cb} + \delta_{il}\delta_{kj}\delta_{ab}\delta_{cd}) \right) \\
&\quad \times M(Q_{ii}Q_{jj} + Q_{ij}Q_{ji}).
\end{aligned} \tag{D26}$$

This neatly reduces to,

$$\mathbb{E}_{U(M)}[D(\mathbf{U}z)^2] = \frac{|z|^4}{M(M+1)} (\mathrm{Tr}(\mathbf{Q}^2) + \mathrm{Tr}(\mathbf{Q})^2) \tag{D27}$$

$$= \frac{|z|^4}{M(M+1)} \mathrm{Tr}(\mathbf{Q}^2), \tag{D28}$$

because the trace of $\mathbf{Q}$ is zero.

#### 4. Distribution of unitary matrix elements

In this section we formally derive the distribution of the unitary-scrambled detection statistic. Although the distribution of the elements of unitary matrices under the Haar measure is likely a standard result in the random matrix literature, we have not been able to find references to it. We first formally derive the result in Sec. D 4 a before we derive the distribution of the detection statistic in Sec. D 4 b.

##### a. Distribution of squared modulus $|U_{jk}|^2$

Let $z = \mathbf{U}\hat{e}$, where $z \in \mathbb{C}^M$, $\mathbf{U} \sim \mathcal{H}U(M)$, and $\hat{e} = (1, 0, 0, \ldots, 0)$ is the first unit basis vector in $\mathbb{C}^M$. This means that $\sum_j |\hat{e}_j|^2 = \sum_j |z_j|^2 = 1$. We are interested in the distribution of $|z_j|^2$ given that $\mathbf{U}$ is Haar distributed. Since $\mathbf{U}$ is Haar distributed, this means that $z$ is distributed uniformly on the unit sphere in $\mathbb{C}^M$. We express each component $z_j$ in squared polar form,

$$z_j = \sqrt{l_j} \exp(J\phi_j), \tag{D29}$$

where $l_j = |z_j|^2$ and $\phi_j$ is the phase. The constraint on the squared modulus of $z$ is thus: $\sum_j l_j = 1$. To find the joint distribution of $(l_1, l_2, \ldots, l_M)$, we calculate the Jacobian of the transformation from $(x, y)$ to $(l, \phi)$. We already know that $z = \mathbf{U}\hat{e}$ is a unitary transform with determinant 1. For the squared polar coordinates we find,

$$\frac{\partial(x_j, y_j)}{\partial(l_j, \phi_j)} = \begin{bmatrix} \frac{\partial x_j}{\partial l_j} & \frac{\partial x_j}{\partial \phi_j} \\ \frac{\partial y_j}{\partial l_j} & \frac{\partial y_j}{\partial \phi_j} \end{bmatrix} \tag{D30}$$

$$= \begin{bmatrix} \frac{1}{2\sqrt{l_j}} \cos\phi_j & -\sqrt{l_j} \sin\phi_j \\ \frac{1}{2\sqrt{l_j}} \sin\phi_j & \sqrt{l_j} \cos\phi_j \end{bmatrix}, \tag{D31}$$

which means that,

$$\left|\det\left(\frac{\partial(x_j, y_j)}{\partial(l_j, \phi_j)}\right)\right| = \frac{1}{2}. \tag{D32}$$

This shows that the Jacobian of our squared polar transformation is invariant under unitary transformations of $z$. Since the Haar measure is invariant under unitary transformations we know that $\mathrm{d}x_j \mathrm{d}y_j$ is invariant under $\mathbf{U}$. And due to the above Jacobian, we therefore have a uniform distribution in $(l_j, \phi_j)$ under the Haar measure. It follows that the joint distribution of $(l_1, l_2, \ldots, l_M)$ is uniform over the $M-1$ dimensional simplex defined by $\sum_j l_j = 1$. The uniform distribution over such a simplex is called the *uniform Dirichlet distribution*. The Dirichlet distribution is a well-known distribution that occurs naturally in many settings, such as the conjugate prior for the multinomial distribution. The uniform Dirichlet distribution, denoted as Dirichlet $(1, 1, \ldots, 1)$, has parameters all equal to 1 for all dimensions.

With this, we have shown that the elements $U_{jk}$ of Haar-distributed unitary matrices $\mathbf{U} \sim \mathcal{H}U(M)$ have as a property that the squared modulus of the elements of the columns of $\mathbf{U}$ are distributed according to a uniform Dirichlet distribution,

$$l = (|U_{j1}|^2, |U_{j2}|^2, \ldots, |U_{jM}|^2), \tag{D33}$$

$$l \sim \mathrm{Dirichlet}(1, 1, \ldots, 1). \tag{D34}$$

This is likely a well-known result in the random matrix literature. However, we have not been able to find a reference that mentions or proves it. Numerical simulations using random draws from the unitary group using SCIPY [60] and NUMPY [61] show our derivation to be correct.

#### b. Distribution of the scrambled detection statistic

The detection statistic has the form (we are omitting tildes and the custom inner product here for clarity),

$$D(z, \theta) = z^\dagger \mathbf{Q} z. \tag{D35}$$

When $z$ is being scrambled with Haar-distributed unitary matrices $\mathbf{U} \sim \mathcal{H}U(M)$, this becomes,

$$D(z, \theta) = z^\dagger \mathbf{U}^\dagger \mathbf{Q} \mathbf{U} z. \tag{D36}$$

As we have seen before, by the definition of the Haar measure, this is equivalent to

$$D(z, \theta) \sim z^\dagger \mathbf{U}^\dagger \mathbf{\Lambda} \mathbf{U} z, \tag{D37}$$

under $\mathbf{U} \sim \mathcal{H}U(M)$, with $\Lambda_{ij} = \delta_{ij}\lambda_i$ the diagonal matrix with eigenvalues $\lambda_i$ of $\mathbf{Q}$ on the diagonal. Writing $z = |z| v$ with $v \in \mathbb{C}^M$ some vector with $|v|^2 = 1$, this becomes,

$$D(z, \theta) \sim |z|^2 v^\dagger \mathbf{U}^\dagger \mathbf{\Lambda} \mathbf{U} v, \tag{D38}$$

$$\sim |z|^2 \hat{e}^\dagger \mathbf{V}^\dagger \mathbf{U}^\dagger \mathbf{\Lambda} \mathbf{U} \mathbf{V} \hat{e}, \tag{D39}$$

$$\sim |z|^2 \hat{e}^\dagger \mathbf{U}^\dagger \mathbf{\Lambda} \mathbf{U} \hat{e}, \tag{D40}$$

where $\hat{e} = (1, 0, \ldots, 0)$ is the first unit basis vector, where $v = \mathbf{V}\hat{e}$ with $\mathbf{V} \in U(M)$. Because $\mathbf{U}\hat{e}$ is just the first row of $\mathbf{U}$, by Eq. (D33) we now have,

$$D(z, \theta) = |z|^2 \sum_j \lambda_j w_j, \tag{D41}$$

$$w \sim \mathrm{Dirichlet}(1, 1, \ldots, 1). \tag{D42}$$

We have now shown that the detection statistic under Haar-distributed unitary transforms is equal to a weighted uniform Dirichlet distribution, where the weights $\lambda_j$ are the eigenvalues of the noise-weighted quadratic filter, and $z$ is the noise-weighted data.

#### c. Scrambling variance through the Dirichlet distribution

Now that we have derived the distribution of the detection statistic under Haar distributed unitary scrambles, we can use the properties of the Dirichlet distribution from Appendix C 2 to rederive the mean and variance of the detection statistic under scrambling. This is a nice check of the results in Appendix D 3. We start with Eq. (D41). The expected value of the detection statistic becomes,

$$\mathbb{E}_{U(M)}[D(z, \theta)] = |z|^2 \mathbb{E}_{U(M)}[w] \sum_j \lambda_j = 0. \tag{D43}$$

Here we have used the fact that the Dirichlet distribution is uniform, so we can pull the expectation over $w_j$ out of the sum, and since the quadratic filter $\mathbf{Q}$ is traceless, the sum over all eigenvalues $\lambda_j$ is zero. This is consistent with what we found in Eq. (66) and Eq. (D21). The variance of the detection statistic can be similarly calculated,

$$\mathbb{E}_{U(M)}[(D(z, \theta))^2] = |z|^4 \mathbb{E}_{U(M)}\left[\left(\sum\nolimits_j w_j \lambda_j\right)^2\right]. \tag{D44}$$

Using Eqs. (C13), (C14) we can rewrite this to be,

$$\mathbb{E}_{U(M)}[(D(z, \theta))^2] = |z|^4 \mathbb{E}_{U(M)}\left[\sum_j w_j^2 \lambda_j^2 + \sum_{j \neq k} w_j w_k \lambda_j \lambda_k\right] = |z|^4 \left(\frac{M-1}{M^2(M+1)} \sum_j \lambda_j^2 - \frac{1}{M^2(M+1)} \sum_{j \neq k} \lambda_j \lambda_k\right). \tag{D45}$$

Now we observe that the quadratic filter $\mathbf{Q}$ is traceless, meaning that the sum of the eigenvalues is zero. From this we deduce that,

$$\left(\sum\nolimits_j \lambda_j\right)^2 = \sum_j \lambda_j^2 + \sum_{j\neq k} \lambda_j \lambda_k = 0 \tag{D46}$$

$$\sum_{j\neq k} \lambda_j \lambda_k = -\sum_j \lambda_j^2. \tag{D47}$$

Substituting this in Eq. (D45), we see that,

$$\mathbb{E}_{U(M)}[(D(z,\theta))^2] = \frac{|z|^4}{M(M+1)} \mathrm{Tr}(\mathbf{Q}^2). \tag{D48}$$

This is exactly as we found in Eq. (D27).